\documentclass{article}

\usepackage{arxiv}

\usepackage[utf8]{inputenc} 
\usepackage[T1]{fontenc}    
\usepackage{hyperref}       
\usepackage{url}            
\usepackage{booktabs}       
\usepackage{amsfonts}       
\usepackage{nicefrac}       
\usepackage{microtype}      
\usepackage{lipsum}
\usepackage{graphicx}
\graphicspath{ {./images/} }

\usepackage{float}
\usepackage{nccmath}
\usepackage[title]{appendix}
\usepackage{nameref}
\usepackage{chngcntr}

\usepackage{amssymb}

\title{A general framework for optimising cost-effectiveness of pandemic response under partial intervention measures}

\author{
 Quang Dang Nguyen \\
  Centre for Complex Systems\\
  Faculty of Engineering\\
  University of Sydney\\
  New South Wales, Australia \\
  \texttt{dang.q.nguyen@sydney.edu.au} \\
   \And
 Mikhail Prokopenko \\
  Centre for Complex Systems\\
  Faculty of Engineering\\
  University of Sydney\\
  New South Wales, Australia \\
  \texttt{mikhail.prokopenko@sydney.edu.au} \\
}

\begin{document}
\maketitle
\begin{abstract}
The COVID-19 pandemic created enormous public health and socioeconomic challenges. The health effects of vaccination and non-pharmaceutical interventions (NPIs) were often contrasted with significant social and economic costs. We describe a general framework aimed to derive adaptive cost-effective interventions, adequate for both recent and emerging pandemic threats. We also quantify the net health benefits and propose a reinforcement learning approach to optimise adaptive NPIs. The approach utilises an agent-based model simulating pandemic responses in Australia, and accounts for a heterogeneous population with variable levels of compliance fluctuating over time and across individuals. Our analysis shows that a significant net health benefit may be attained by adaptive NPIs formed by partial social distancing measures, coupled with moderate levels of the society’s willingness to pay for health gains (health losses averted). We demonstrate that a socially acceptable balance between health effects and incurred economic costs is achievable over a long term, despite possible early setbacks.
\end{abstract}


\section{Introduction}
\label{sec:main}

The COVID-19 pandemic has generated enormous health, economic and social costs, causing a significant loss of life, adversely affecting the population health, and creating a substantial shock to national and global economies. The  pandemic dramatically reduced life expectancy and increased premature mortality~\cite{andrasfay2021reductions,islam2021effects,aburto2022quantifying}, often impairing the capacity of healthcare systems to deal with the crisis~\cite{remuzzi2020covid,miller2020disease}. In parallel, various supply chains, labour and equity markets worldwide, and entire sectors of economy, such as tourism, energy and finance sectors, have also suffered very substantial and cascading impacts \cite{inoue_propagation_2020,mayhew_covid-19_2020,del_rio-chanona_supply_2020,bruce_financial_2022}.
These non-trivial challenges created a major need for appropriate and sustainable pandemic responses capable of balancing both health and socioeconomic consequences: a problem which continues to evade simple solutions. On the one hand, it proved to be difficult to objectively model and quantify the health and economic costs in comparative terms~\cite{anderson2020developing,padhan2021economics}. On the other hand, contrasting the health and socioeconomic impacts was often dependent on fairly subjective perspectives of policy- and decision-makers~\cite{maor2020when,lee2020policy,norheim2021difficult}, as well as cultural differences, political influences, and other factors affecting trust in governments and health authorities~\cite{seale2020covid,trent2021trust}.  Typically, attempts to ``optimise'' multiple objectives were carried out under severe pressure, failing to flatten epidemic curves and prevent an economic downturn.  Consequently, society-wide responses were prone to conflicting influences, including (geo-)political, economic, social and behavioural factors, and produced sub-optimal or ill-timed intervention measures.

Initial responses to the unfolding pandemic, dominated by an abundance of caution and employing mostly non-pharmaceutical interventions (NPIs), have varied in their scope and effectiveness~\cite{walker2020impact,flaxman2020estimating,amtrac_19,mamelund_social_2021,solomon_adherence_2022}. Once vaccines became broadly available, the intervention focus has changed to mass vaccination campaigns which have also shown a varying degree of success, reducing the pandemic impact and allowing many affected societies and economies to moderately recover~\cite{moore2021vaccination,harris2021impact,viana2021controlling,mass_vaccination_and_lockdown}. Nevertheless, subsequent pandemic waves generated by emerging viral variants of concern, such as the Delta and Omicron variants of SARS-CoV-2, continued to severely disrupt individual lives and national economies~\cite{campbell2021increased,milne2022mitigating,karim2021omicron}. It is likely that the ongoing spread and evolution of the SARS-CoV-2 virus will continue to affect the recovery efforts, with periods of relative calm interleaved with renewed outbreaks and pandemic waves. Possible oscillatory pandemic dynamics can be expected even after a transition to endemicity, with only transient herd immunity developing in the near to mid-term, without long-lasting transmission-blocking immunity~\cite{antia2021transition}. Moreso, a transition to endemicity is dependent on the interplay of human behaviour, demographics, susceptibility, immunity, and emergence of new viral variants~\cite{katzourakis2022COVID}.


It is increasingly evident that maintaining a strict intervention policy is hardly possible over a long time, especially if such a policy involves persistent social distancing and stay-at-home orders. Given the imperative to balance population health against unacceptable or severe socio-economic impacts, in presence of emerging outbreaks, 
there is a clear need for a more refined approach centred on adaptive, contextual and cost-effective interventions. A successful approach should not only offer a way to reconcile health and socio-economic perspective, but also utilise an unbiased methodology for a search of optimal or near optimal policies, non influenced by modelling or policy-making preferences.


This study addresses several of these challenges. Firstly, we quantify the cost-effectiveness of intervention measures using the Net Health Benefit (NHB) approach which balances \emph{both health and economic costs}. In doing so, we formulate a search space for suitable interventions in terms of two thresholds. The first one is the ``willingness to pay'' (WTP) defined as an acceptable threshold that can be paid for an additional unit of health benefit, such as the disability-adjusted life years (DALY) averted. The second threshold is the level of maximal compliance with the social distancing (SD) measures that may be imposed on the population. Varying the WTP per DALY averted for different SD levels allows us to systematically explore the policy search space with respect to the resultant health benefits. 

Secondly, in modelling the transmission and control of the COVID-19 pandemic we adopt an agent-based model (ABM) approach, described in Methods and Appendix \ref{sec:SM_ABM}: \nameref{sec:SM_ABM}. Each individual is represented by a computational agent stochastically generated based on relevant demographic (census) data. These agents are simulated over time with respect to their interactions in different social contexts (e.g., residential, workplace, educational), probabilistically transmitting infection, getting ill, recovering from the disease, and so on~\cite{amtrac_19,chang2022simulating}. This high-resolution approach allows us to capture not only the \emph{population heterogeneity}, but also the \emph{fluctuating compliance} with social distancing that may vary (i) across the agents, and (ii) over time. In each simulated scenario, we do not assume homogeneous or persistent adherence of individuals to a fixed SD level, instead limiting the fraction of compliant individuals, dynamically selected at each time point, by a maximal level. 
Thus, the model can evaluate   \emph{partial} interventions shaped by complex SD behaviours fluctuating between the strongest possible commitment ($SD = SD_{max}$) and extreme fatigue ($SD = 0$), heterogeneously and dynamically distributed across the population.

Following~\cite{chang2022simulating}, social distancing is defined holistically, comprising various behavioural changes aimed to reduce the intensity of individual interactions during a given restriction period. These changes typically include stay-at-home restrictions, travel reduction, as well as physical distancing, mask wearing, and other measures. Dependent on the social context, each compliant agent may reduce the intensity of interactions with their household members, neighbours and coworkers/classmates. A typical scenario also assumes other NPIs, such as case isolation and home quarantine, as well as a \emph{partial} mass-vaccination coverage {affecting a proportion of the population (see Methods and Appendix \ref{sec:SM_ABM}: \nameref{sec:SM_Vaccination}). In general, vaccination campaigns can be either pre-emptive (vaccination between outbreaks) or reactive (vaccination during an outbreak). We consider pre-emptive vaccination campaigns in relation to future outbreaks of the COVID-19. In other words, we assume that a vaccination campaign, covering a significant fraction of the population (e.g., 85\%), is carried out before such an outbreak caused by an emerging variant of concern. Before exploring the search space of interventions, we calibrated and validated the ABM using a case study: an outbreak of the Delta variant in New South Wales, Australia, during June--November 2021 (see Appendix \ref{sec:SM_ABM}: \nameref{sec:SM_Validation}).

Finally, the study utilises a reinforcement learning (RL) algorithm exploring the search space of feasible cost-effective NPIs. Rather than formulating an NPI in advance, specifying exact SD compliance levels for each time interval, the RL algorithm constructs possible interventions dynamically. This is achieved by selecting possible future SD actions based on relative success of prior simulations, i.e., using a ``reward'' signal. These rewards may reinforce or weaken the selection probability of the corresponding SD actions, so that better policies may emerge after a sufficiently long learning period.  In order to ensure feasibility, the actions are selected at some realistic decision points (e.g., weekly).  The interventions that outperform other candidates in terms of cost-effectiveness, as measured by the NHB --- that is, balancing both health and economic costs --- contribute to further simulations, improving the NHB over the learning process. Crucially, the RL-based approach eliminates subjectivity in selecting feasible NPIs, by removing bias towards several frequently considered preferences for ``short and snap'' lockdowns~\cite{blakely2020probability}, mandatory large-scale social distancing campaigns~\cite{walker2020impact,flaxman2020estimating}, or loose restrictions in style of de-facto ``herd immunity'' approaches~\cite{claeson2021covid,mishra2021comparing}. In principle, resultant interventions produced by the RL algorithm may have different temporal profiles unencumbered by such subjective choices, as long as the policy changes are deemed feasible and the outcomes are superior. 

Our comparative analysis shows that it is possible to generate a significant net health benefit by a feasible pattern of partial social distancing measures. These temporal patterns, produced by an unbiased machine learning approach, adapt to different values of the society's willingness to pay for a single lost (disability-adjusted) year of life. A resultant profile typically starts with near-maximal levels of adherence to social distancing ($SD \approx SD_{max}$) and allows for a significant relaxation of social distancing to much lower levels ($SD \lessapprox 0.2$) after several weeks. While the period of higher commitment is dependent on the WTP threshold, this dependence does not preclude viable SD interventions progressing even for relatively low values of the threshold $SD_{max}$. However, with higher $SD_{max}$ thresholds, the relaxation of strict compliance measures becomes viable sooner. Interestingly, the net health benefit produced by mid-level compliance with NPIs carried out under a moderate willingness-to-pay setting is commensurate with the benefits yielded by higher $SD_{max}$ and WTP. 

In adaptive strategies, the level of SD imposed on the population in a given week changes in response to the current pandemic state, and thus depends on previous actions in context of the accepted WTP. Importantly, adaptive NPIs outperform possible alternatives, including fixed, random or zero social distancing, across the entire range of considered WTP and $SD_{max}$ thresholds. We demonstrate that the higher cumulative NHB attained by adaptive NPIs non-linearly balances the incurred economic costs and sustained health losses, achieving longer-term advantages despite possible early setbacks. Overall, these findings suggest that the choice between ``health'' and ``economy'' is a false choice, and an adaptive policy may achieve a socially acceptable balance even under significant constraints.

\section{Results}

\label{sec:results}

Using our agent-based model and reinforcement learning algorithm, we investigated three settings of maximal compliance with social distancing ($SD_{max} = 0.3$, $SD_{max} = 0.5$, and $SD_{max} = 0.7$, in other words, the maximal fraction of compliant population set at 30\%, 50\% and 70\% respectively). For each of these levels, we varied the  ``willingness to pay'' (WTP) across three thresholds: \$10K per DALY, \$50K per DALY and \$100K per DALY.  Each (SD, WTP) setting was evaluated in terms of the Net Health Benefit (NHB), generated by the adaptive NPIs learned by the RL algorithm over 19 simulated weeks of the disease spread, and further analysed with respect to  two separate NHB components: the economic costs and health effects. Each simulation run was carried out for a typical pandemic scenario developing in an Australian town (see Methods and Appendix \ref{sec:SM_ABM}: \nameref{sec:SM_ABM}), with the economic costs adjusted in proportion to the population size, and DALY losses computed from the estimates of incidence and fatalities generated by multiple ABM simulations.

\textbf{Dynamics of adaptive NPIs.}
Figure~\ref{fig:eval_policies_SD30_SD70} contrasts two $SD_{max}$ settings, 30\% and 70\%, across three considered WTP levels, tracing the level of compliance with social distancing over time. A similar comparison between $SD_{max}$ set at 50\% and 70\% is presented in Appendix (Fig.~\ref{fig:eval_policies_SD50_SD70}).  
All adaptive NPIs, triggered when the number of detected infections exceeds a threshold (invariably, during the first week), begin at the maximal considered SD level, i.e., $SD \approx SD_{max}$.  In general, the SD level reduces over time. While this reduction is gradual and almost linear for the low-commitment setting, $SD_{max} = 0.3$ (Fig.~\ref{fig:eval_policies_SD30_SD70}: left panels), it is more abrupt and non-linear for the high-commitment setting, $SD_{max} = 0.7$, with the decline evident after just a few weeks (Fig.~\ref{fig:eval_policies_SD30_SD70}: right panels). In other words, higher initial levels of SD compliance allow for NPIs with shorter periods of stricter stay-at-home orders.  

The differences across the patterns produced by distinct $SD_{max}$ settings strongly suggests that an early suppression of outbreaks helps to increase the NHB: an outcome achieved even with  SD levels declining over time. Importantly, a comparison of the top, middle and bottom panels of Fig.~\ref{fig:eval_policies_SD30_SD70} shows that the SD levels attained at the end of simulations depend on the WTP threshold, with the higher WTP values generating higher convergent SD commitments across all considered levels of $SD_{max}$. For example, for WTP set at \$10K per DALY, the NPIs beginning at $SD_{max} = 0.7$ (top-right panel) converge to merely 5\% of SD compliance, while for WTP set at \$100K per DALY, the NPIs beginning at $SD_{max} = 0.7$ (bottom-right panel) remain above 20\% of SD compliance at the end. A similar tendency is observed for low maximal commitment $SD_{max} = 0.3$ as well, with convergent SD levels differentiated between 10\% of SD compliance (top-left panel) and slightly below 20\% of SD compliance (bottom-left panel). This highlights the role of WTP threshold in reducing the DALY losses while minimising the corresponding economic costs.

The reported observations are robust, as indicated by boxplots shown in Fig.~\ref{fig:eval_policies_SD30_SD70}, with the majority (at least 50\% shown within the box body) of the simulations following the described patterns. 

\begin{figure}
    \centering
    \includegraphics[width=\textwidth]{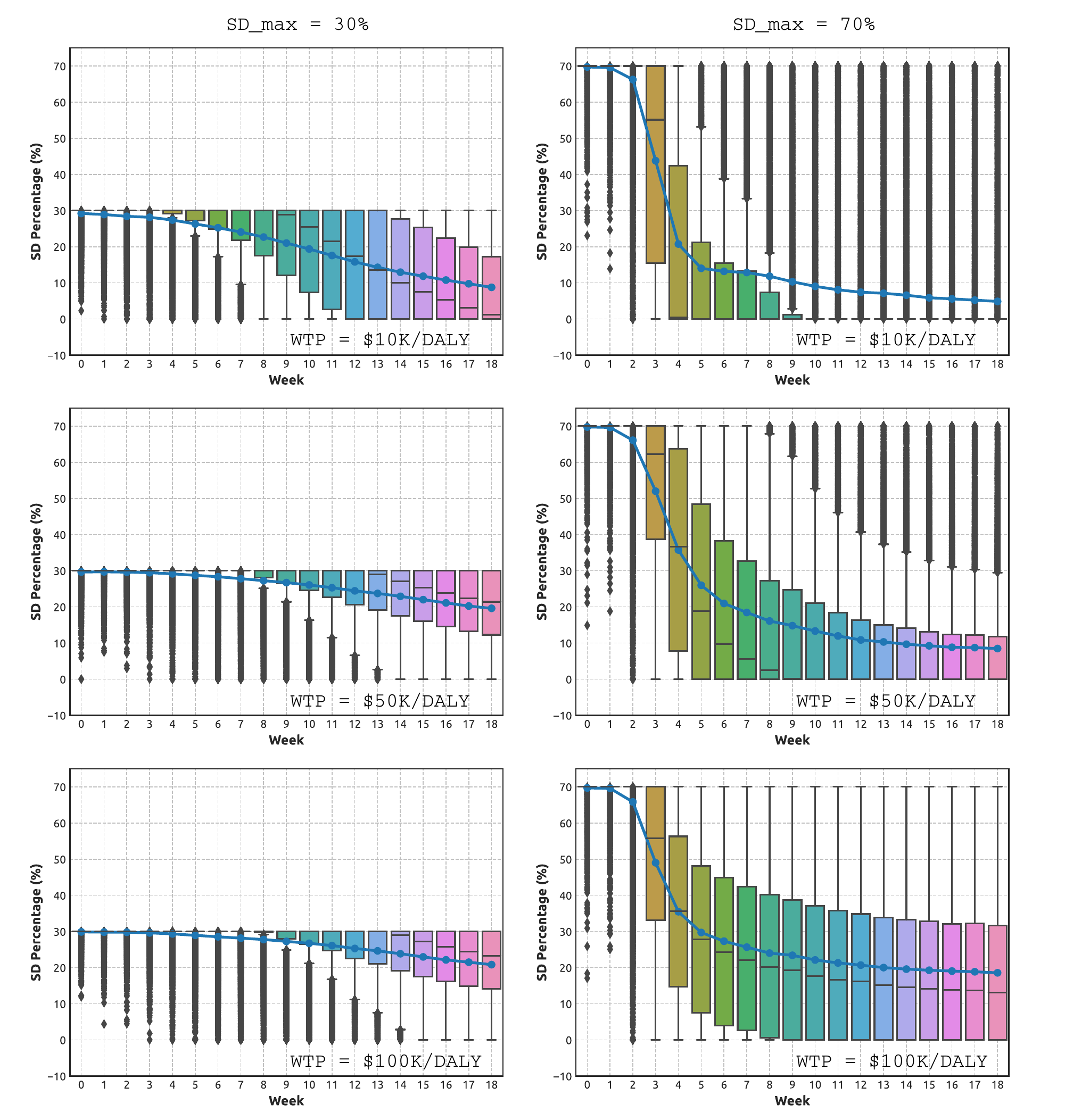}
    \caption{Adaptive NPIs, learned under different combinations of maximal SD levels $SD_{max}$ and WTP, over more than 14,000 simulations. Left: $SD_{max} = 0.3$. Right: $SD_{max} = 0.7$. Top: WTP is set at \$10K per DALY. Middle: WTP is set at \$50K per DALY. Bottom: WTP is set at \$100K per DALY. Boxplots show the distribution of data over the quartiles, with box body capturing the mid-50\% of the distribution. The curves shown with blue colour trace the mean values of the SD levels attained in each week. Outliers are shown in black.}
    \label{fig:eval_policies_SD30_SD70}
\end{figure}


\textbf{Dynamics of Net Health Benefit.}
Having examined dynamics of the adaptive NPIs produced by reinforcement learning, we turn our attention to dynamics of the corresponding NHB generated by these NPIs. Figure~\ref{fig:cumulative_rewards_SD30_SD70} compares adaptive NPIs against several competing intervention approaches, across the considered threshold levels of $SD_{max}$ and WTP (see also Appendix (Fig.~\ref{fig:cumulative_rewards_SD50_SD70})). The alternatives include: (i) fixed SD levels, when SD is set at $SD_{max}$ for the entire duration of the simulation; (ii) random SD levels, randomly fluctuating between 0 and $SD_{max}$ during the simulation; and (iii) no social distancing (zero SD).  The adaptive NPIs outperform all the contenders, achieving a higher cumulative NHB for the majority of (SD, WTP) settings, and the same NHB for two settings. 

Importantly, dynamics of the NHB over time are non-linear, allowing us to contrast short-term and long-term advantages of the adopted measures. The adaptive NPIs generate a superior cumulative NHB despite some initial losses or slow gains. In particular, the low WTP threshold of \$10K per DALY, traced in the top panels of Fig.~\ref{fig:cumulative_rewards_SD30_SD70}, produces negative NHB for a number of weeks: three weeks for $SD_{max} = 0.3$ and four weeks for $SD_{max} = 0.7$, followed by positive cumulative NHB during the rest of simulation. For higher WTP thresholds, \$50K per DALY and \$100K per DALY (shown in the middle and bottom panels), the NHB remains positive during an initial period, growing relatively slowly, followed by a more rapid NHB increase plateauing towards the end.  

In some cases, during the initial period (i.e., the first few weeks), the cumulative NHB produced by the adaptive interventions is smaller than the NHB produced by the alternatives, but this is invariably replaced by higher NHB gains generated by adaptive interventions over a longer term. Thus, the intervention window offered by the initial period is crucial for an effective control of the pandemic, regardless of the society's willingness to pay per DALY loss averted. This observation aligns with the other studies advocating early SD measures aimed to prevent escalation of outbreaks \cite{demirguc-kunt_sooner_2020,kompas_health_2021,binny_early_2021}, indicating that early short-term sacrifices yield longer-term benefits. The balance between short- and long-term advantages is most striking for the low WTP threshold of \$10K per DALY, while the higher WTP thresholds extract the NHB gains almost immediately.

The approach with zero SD intervention is obviously not a serious contender, but offers a useful baseline in terms of delineating these short- and long-term advantages. Specifically, the time point when the cumulative NHB of an adaptive NPI exceeds the corresponding level for the zero SD intervention marks the point when the adaptive intervention starts to generate a longer-term benefit. 

A fixed-SD intervention presents stronger competition, and achieves the same NHB as the adaptive policy in two cases: low maximal SD commitment ($SD_{max} = 0.3$) for middle and high WTP thresholds: \$50K per DALY and \$100K per DALY (shown in middle-left and bottom-left panels). These two outcomes suggest that when the the society's willingness to pay per DALY averted is high, the low maximal SD commitment, such as $SD_{max} = 0.3$, constraints the scope for adaptive interventions. Nevertheless, even under this constraint, the proposed reinforcement learning approach is successful in finding adequate interventions which are as effective as their fixed SD counterparts while being less stressful for the society. 

Continuing with Fig.~\ref{fig:cumulative_rewards_SD30_SD70}, we also point out that the fixed SD approach essentially fails in some settings (e.g., high maximal SD commitment, $SD_{max} = 0.7$, for low WTP threshold: \$10K per DALY), and sometimes, performs as poorly as the random SD approach (e.g., high maximal SD commitment, $SD_{max} = 0.7$, for middle WTP threshold: \$50K per DALY). The lower WTP thresholds bias the NHB to weigh the economic costs of intervention higher than the health effects (averted DALY losses). Evidently, the fixed SD intervention cannot cope well with this bias, especially when there is a scope for higher compliance with the stay-at-home restrictions. In contrast, the adaptive NPIs perform very well under the higher maximal SD compliance, such as $SD_{max} = 0.7$, extracting a higher NHB than the alternatives, with the relative gains being much higher for the smallest considered WTP threshold.

Finally, a comparison between the left and right panels of Fig.~\ref{fig:cumulative_rewards_SD30_SD70} shows that the higher levels of $SD_{max}$ allow the interventions to attain the higher NHB gains, regardless of the WTP threshold, and this potential is fully realised by the adaptive NPIs.

\begin{figure}[H]
    \centering
    \includegraphics[width=\textwidth]{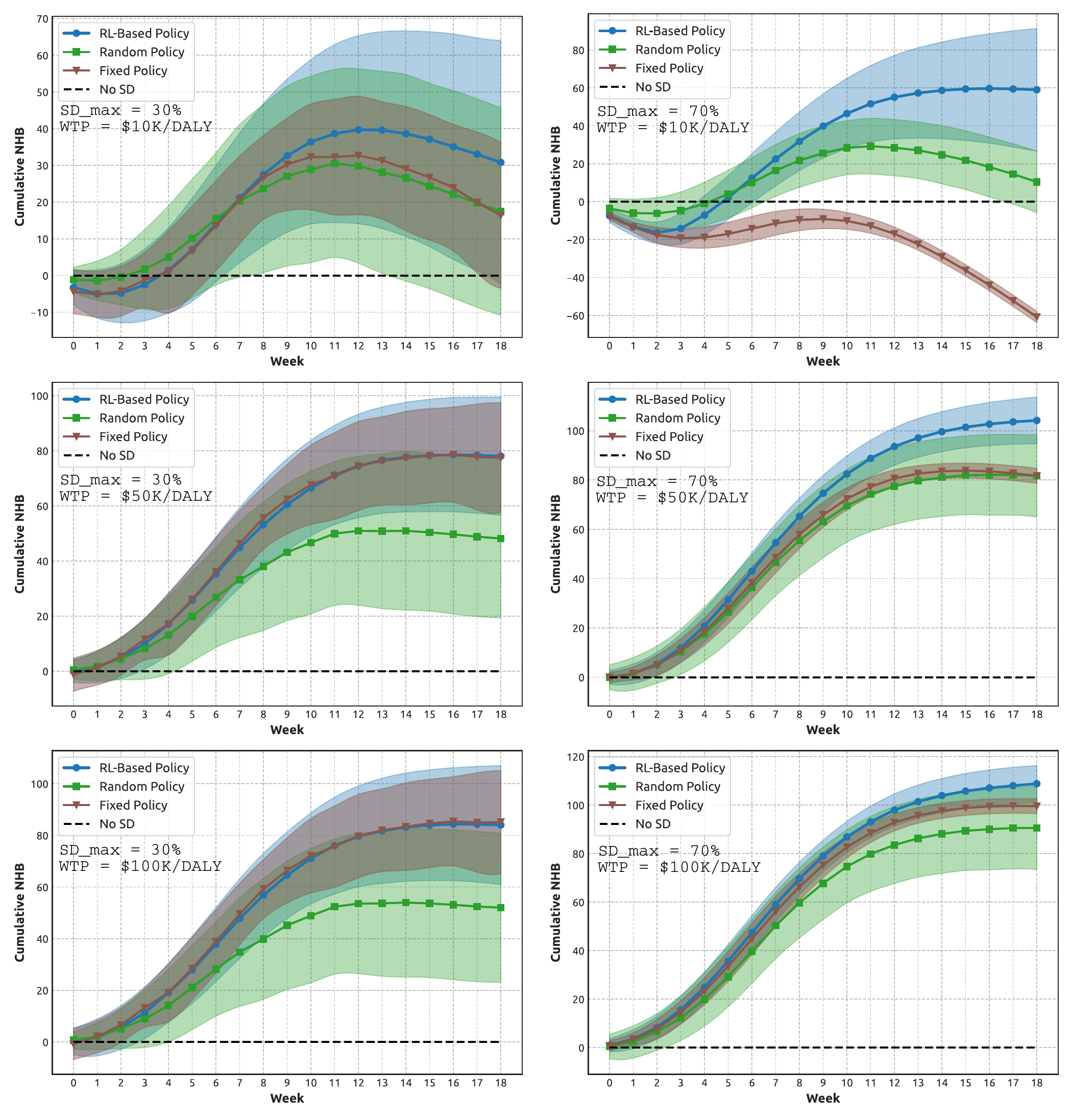}
    \caption{A comparison of cumulative Net health benefit (NHB) generated by the adaptive NPIs, fixed SD NPIs, random SD policies, and zero SD policies.  Left: $SD_{max} = 0.3$. Right: $SD_{max} = 0.7$. Top: WTP is set at \$10K per DALY. Middle: WTP is set at \$50K per DALY. Bottom: WTP is set at \$100K per DALY. Shaded areas show standard deviation.} 
    \label{fig:cumulative_rewards_SD30_SD70}
\end{figure}


Figure~\ref{fig:nhb_heatmap} summaries the average cumulative NHB generated by adaptive NPIs trained under different configurations of the WTP thresholds and the maximal SD compliance levels $SD_{max}$. The heatmap exhibits a clear NHB gradient towards the higher WTP and $SD_{max}$. 
However, as $SD_{max}$ increases (left to right), the relative NHB gains diminish. Hence, stricter intervention measures offer a relatively smaller gains in the health benefit, especially for higher WTP thresholds. Similarly, as WTP threshold increases (top to bottom), there are marginal gains in the corresponding NHB. This indicates that a progressively higher WTP threshold does not necessarily translate into a proportionally smaller DALY losses. Arguably, a mid-level WTP threshold coupled with mid-level compliance with the stay-at-home restrictions attains the net health benefit comparable with the higher WTP and $SD_{max}$.

\begin{figure}[H]
    \centering
    \includegraphics[scale=0.625]{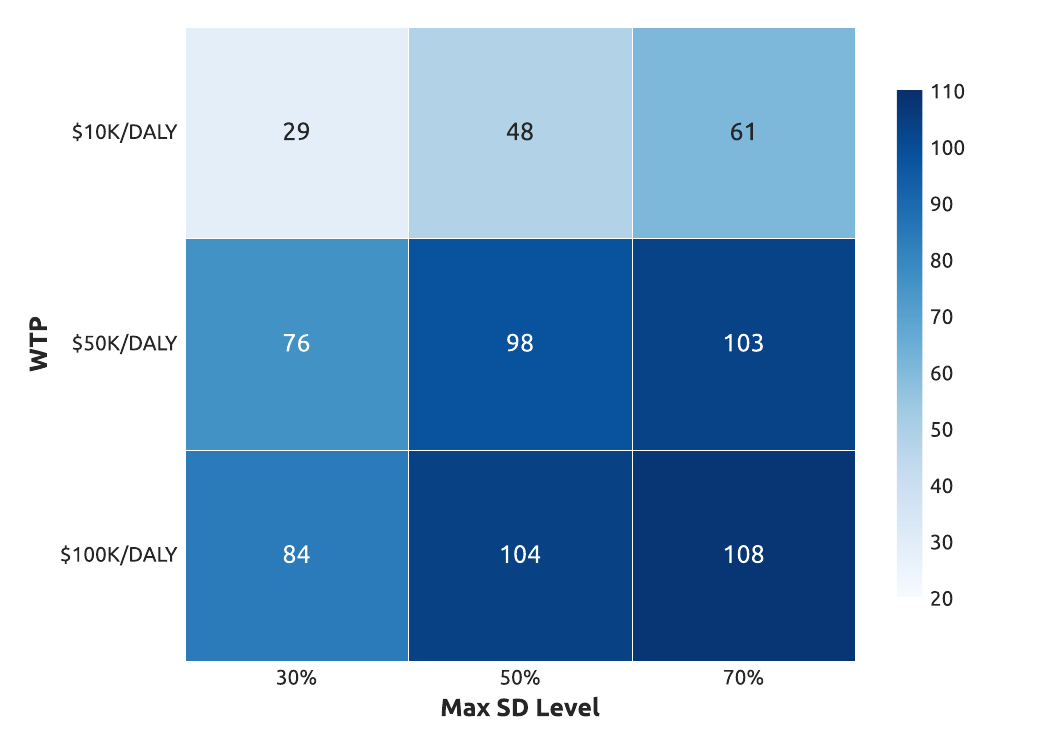}
    \caption{Cumulative Net Health Benefit (NHB) generated by adaptive NPIs, evaluated in the space of two threshold parameters: maximal SD level $SD_{max}$ and willingness-to-pay (WTP) threshold. NHB values shown in each cell are the mean values produced over approximately 14,000 simulations.}
    \label{fig:nhb_heatmap}
\end{figure}


\textbf{Composition of the Net Health Benefit (NHB).}
The NHB is shaped by two components: economic costs and health effects (i.e., averted DALY losses), and we now analyse this composition explicitly. 
Figure \ref{fig:economic_cost_and_health_loss} compares the economic costs and health effects (averted DALY losses) accumulated over the studied period in response to different NPIs. As above, the NPIs include adaptive SD learned across three WTP thresholds (\$10K per DALY, \$50K per DALY and \$100K per DALY), as well as three alternatives: random SD, fixed SD, and zero SD. The analysis is carried our for three maximal levels of compliance: $SD_{max} \in \{0.3, 0.5, 0.7\}$, see also Appendix (Fig.~\ref{fig:economic_cost_and_health_effect_SD50_SD70}).

The zero SD interventions provide a baseline, showing zero economic costs and zero averted DALY losses (i.e., substantial health losses). The fixed SD interventions provide an opposite baseline, showing linearly growing economic costs (with the slope determined by the constant costs incurred per week in order to maintain the maximal SD level), and significant cumulative health effects in terms of averted DALY losses. The economic costs of the fixed SD interventions are the highest among alternatives, allowing to generate the highest associated health effects. Random SD interventions produce the economic costs and health effects between these two opposite baselines. 

The adaptive SD interventions outperform its random alternatives, producing outcomes which are closer to one of the opposite baselines, across all considered levels of $SD_{max}$. In particular, despite incurring lower costs than the fixed and random SD interventions, the adaptive NPIs approach the maximal health effects generated by the SD interventions fixed at $SD_{max}$. In other words, the economic costs of adaptive NPIs markedly slow their growth over time, reaching the levels significantly below the costs of the fixed SD interventions, while the corresponding averted DALY losses are almost as high as the DALY losses averted by the fixed SD interventions. In summary, the adaptive NPIs achieve a beneficial trade-off between the economic costs and health effects, demonstrating long-term sustainability.  As expected, the NPIs derived  using the lower WTP threshold (\$10K per DALY) yields lower economic costs, but generates worse health effects (that is, averts fewer DALY losses). Conversely, the NPIs derived using the higher WTP threshold (\$100K per DALY) leads to higher economic costs, but achieves better health effects (averts more DALY losses).

\begin{figure}[H]
    \centering
    \includegraphics[width=\textwidth]{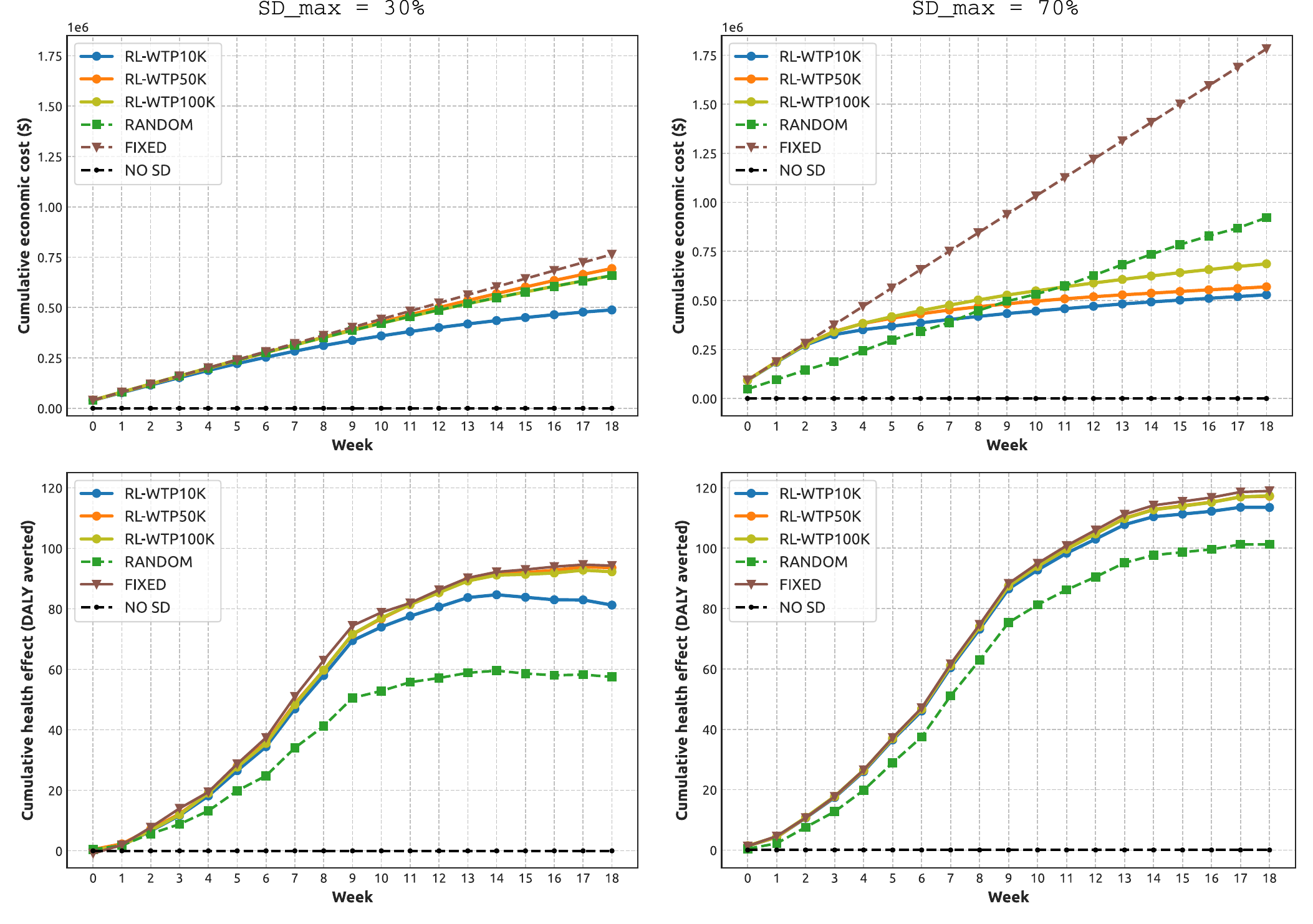}
    \caption{Components of the Net Health Benefit (NHB): mean values of cumulative economic costs (dollars) and cumulative health effect (DALY averted), shown for different NPIs: adaptive SD with three WTP thresholds (\$10K per DALY, \$50K per DALY and \$100K per DALY), random SD, fixed SD, and zero SD.  Left: the maximal SD level $SD_{max}$ is set at 30\%. Right: the maximal SD level $SD_{max}$ is set at 70\%.}
    \label{fig:economic_cost_and_health_loss}
\end{figure}

Finally, we consider a phase diagram of the Net Health Benefit dynamics with respect to two cumulative NHB components: the health effects in terms of DALY losses averted by adaptive NPIs and the economic costs. The diagram, shown in Fig.~\ref{fig:nhb_scatter}, displays results of the adaptive NPIs carried out with different thresholds for WTP and maximal SD compliance $SD_{max}$. It reveals that the search-space sampled by the \emph{optimised} adaptive NPIs is well-structured, with the areas corresponding to different values of $SD_{max}$ clearly delineated. In particular, the area formed by NPIs operating under $SD_{max} = 0.3$ covers the region with low economic costs and low health effects. In contrast, the area formed by NPIs operating under $SD_{max} = 0.7$ covers the region with medium-to-high economic costs and relatively high health effects, but this region has a complex narrow shape, highlighting difficulties of exploring the search-space. The area formed by NPIs operating under medium maximal compliance, $SD_{max} = 0.5$, is large and well-shaped, including the region across an almost entire scale of economic costs and medium-to-high health effects. The most attractive part of all three regions (with low economic costs and high health effects) is reachable in principle but occupies a narrow tip of the attained search-space, with the search being challenged more for higher WTP thresholds.

\begin{figure}[H]
    \centering
    \includegraphics[width=\textwidth]{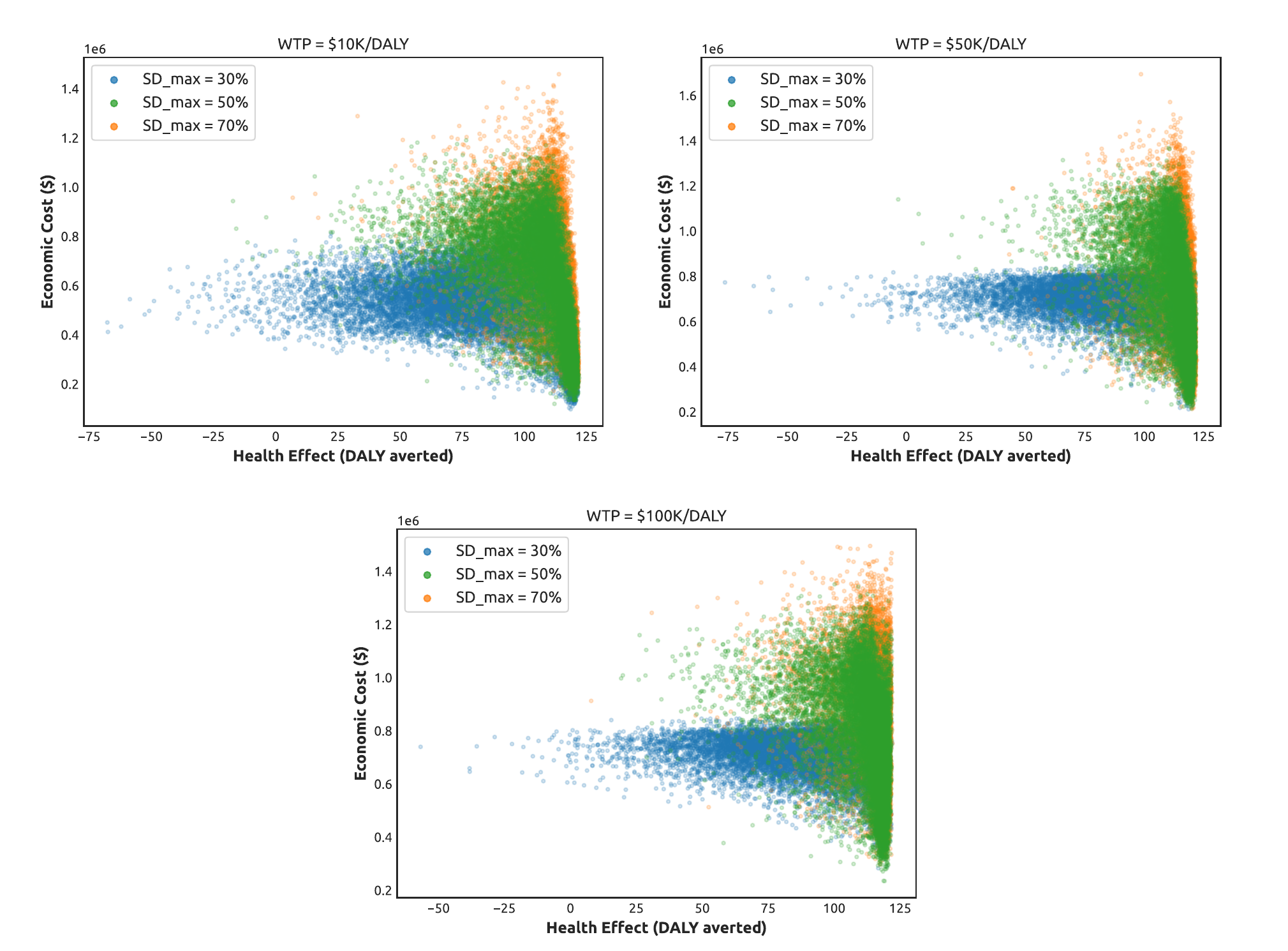}
    \caption{Colour plots of the Net Health Benefit (NHB) as a function of cumulative health gains (horizontal axis) and cumulative economic costs (vertical axis). The cumulative NHB is shown for more than 1000 simulations, carried out for different maximal SD levels: $SD_{max} = 0.3$ (blue), $SD_{max} = 0.5$ (green), and $SD_{max} = 0.7$ (orange). The colour plots are shown for adaptive NPIs with three WTP thresholds: \$10K per DALY, \$50K per DALY and \$100K per DALY. }
    \label{fig:nhb_scatter}
\end{figure}

\section{Discussion}
\label{sec:discussion}
Continuing efforts to control the COVID-19 pandemic typically combine mass vaccination campaigns and diverse non-pharmaceutical interventions (NPIs)~\cite{moore2021vaccination,harris2021impact,viana2021controlling,chang2022simulating}. In general, the socio-economic costs of the imposed NPIs are significant due to scarcity of resources, unequal wealth distribution across different socio-economic profiles, and wavering social acceptance of NPI restrictions driven by the fatigue with lockdowns and other measures. Not surprisingly, the task of finding a cost-effective allocation of finite resources generated an active research effort~\cite{higher_SD_better_icer,healthcare_cost_n_benefits,neumann_systematic_2016}. This task is likely to remain an important societal challenge, exacerbated by continuously evolving variants of the novel coronavirus which reduce the chances of completely eradicating the COVID-19 in the near future.

Traditional approaches quantifying the cost-effectiveness of interventions, such as the methods based on incremental cost-effective ratio (ICER)~\cite{icer_1977, icer_1988}, have encountered some limitations in determining and interpreting cost-effectiveness outcomes, carrying out sensitivity or scenario analysis, ranking strategies, as well as in addressing equity concerns
~\cite{health_econometrics, paulden_why_abandon_icer_2020}. In particular, it was argued that ``observing change in the ICER does not necessarily imply that a strategy is more or less cost effective than in the reference-case analysis'' \cite{paulden_why_abandon_icer_2020,paulden_calculating_2020}. More recent approaches measuring cost-effectiveness utilised the net-benefit analysis. In particular, the concept of Net Health Benefit (NHB) was introduced to formalise the cost-effectiveness problem, linearising the balance between the economic costs and health effects~\cite{net_health_benefit}. Formally, the NHB quantifies the difference between the health effects of two interventions: the one which is being evaluated and the baseline which is not subject to evaluation, given the corresponding cost incurred at a pre-defined cost-effectiveness threshold~\cite{WHO_undertaking_2003}. Similarly, the Net Monetary Benefit (NMB) was defined as a reformulation of the NHB~\cite{net_monetary_benefit}. Both NHB and NMB were shown to offer advantages in building regression-based models used in economic evaluations~\cite{health_econometrics}. Over the last years, the ICER-based and the net-benefit-based methods have been used to evaluate different intervention approaches aimed to control the spread of COVID-19~\cite{icer_south_africa,higher_SD_better_icer,healthcare_cost_n_benefits,Blakely_2020,NMB_physical_distancing}.

In this work, we proposed an approach to quantify the cost-effectiveness of complex intervention measures using the NHB method accounting for both economic costs and health effects. In doing so, we simulated the COVID-19 pandemic scenarios unfolding within a representative heterogeneous population. Using an agent-based model (ABM), we compared various intervention measures, including adaptive NPIs derived by a reinforcement learning (RL) algorithm employed to maximise the NHB. The search-space for the adaptive NPIs was formed by two thresholds: (i) the ``willingness to pay'' (WTP) per unit of the disability-adjusted life years (DALY), and (ii) the level of maximal compliance with the social distancing measures $SD_{max}$. In choosing the three representative levels of WTP per DALY that we considered as thresholds, we followed several prior studies (See Methods: Willingness to pay).

In recent past, several RL methods have been applied to obtain optimal NPIs in context of the COVID-19, including RL-based training of a three-level lockdown policy for a simulated population of ten thousand people~\cite{sci_report_2020}, which was shown to outperform age-based lockdowns~\cite{age_based_lockdown} and n-work-m-lockdowns~\cite{n_work_m_lockdown}. An RL meta-agent-based intervention was derived as a school closure strategy in Great Britain~\cite{gb_school_closure}, while a multi-agent approach to optimise social distancing was evaluated across India~\cite{india_lockdown}. 


The approach presented in our study extends the state-of-the-art in three ways. Firstly, we demonstrate that an NHB-based approach can deliver a sustainable, adaptive, and contextual cost-effective SD intervention policy. The WTP threshold balances economic costs and health effects, contextualising the NHB given the society's willingness to pay for DALY losses averted. Put simply, the higher the WTP, the more weight is placed on the potential health gains achieved by the incurred economic costs.
Given the WTP threshold, the adaptive interventions carried out under an $SD_{max}$ limit are shown to be sustainable over time, reconciling both competing objectives in the long run. This indicates that these objectives do not have to be directly contrasted and various trade-offs are possible in the (WTP, SD) space. 
Importantly, these trade-offs are achievable for medium settings of $SD_{max}$, such as $SD_{max} = 0.5$, and WTP, e.g., WTP = \$50K per DALY.  This shows that future outbreaks can be controlled reasonably well even when the population response to NPIs is partial and fluctuating over time. Such partial response at a medium level of compliance may generate the NHB commensurate with the benefits of a higher but more demanding $SD_{max}$ limit.
Crucially, these trade-offs are realised over a period of several months, and there are clear differences in the cumulative NHB generated during a short- and a long-term.
Our results clearly show that a preference of ``economy'' over ``health'' does not bring long-lasting benefits, and short-term gains quickly dissipate over time. This highlights the need to take a long-term perspective in evaluating net health benefit, considering a sufficiently long response horizon.

Secondly, we capture the heterogeneity of human responses within the fluctuating fraction of compliant population. The ABM simulation not only represents the population heterogeneity, but also accounts for the stochasticity of individual choices with respect to compliance with the SD restrictions at any given time. Thus, the adaptive NPIs learned by the RL algorithm are broadly applicable to populations characterised by diverse socio-economic profiles.

Thirdly, the proposed method producing adaptive NPIs significantly reduces, and ideally removes, the subjectivity and bias which are often present in the public health decision-making. The learning algorithm considers a sufficiently long time horizon without setting any preferences or assuming particular patterns for interventions. Given the high temporal resolution of decision points (weekly) and the continuous range of the compliance levels (between zero and $SD_{max}$), the learned SD profiles are relatively unconstrained, and yet exhibit smooth trajectories, without abrupt changes. The approach also allows for a principled comparative analysis between the adaptive NPIs and their alternatives, such as fixed, random or zero SD interventions.  

In our approach, the decisions about compliance are not taken by the agents. Instead, the interventions assume that a fraction of the population is compliant with the level imposed by a centralised decision-maker. At each simulated day, we assign the compliant agents differently, so that the distribution  within the population varies across time.  Yet the agents do not individually decide on compliance based on personal risk-benefit considerations.  An important extension of this framework would include behavioural factors which are highly influenced by personal perceptions of risk \cite{bedson_review_2021,piraveenan_optimal_2021,chang_effects_2019,chang_game_2020}, as well as peer group pressure \cite{fournet_epidemic_2016,wang_suppressing_2016} and social media messaging \cite{granell_competing_2014,wang_co-evolution_2020,alvarez-zuzek_epidemic_2017}. Incorporating opinion dynamics and risk-benefit analysis within an agent-based model continues to be a subject of future research.

The limitations of the study include the relatively small size of the simulated population (a typical town). The ABM itself was calibrated for a much larger population size, i.e., across millions of agents simulated at the state level (New South Wales). However, learning adaptive interventions at high temporal and SD-level resolutions demanded the reduction in the size of simulated population. In turn, we needed to proportionally reduce the economic costs incurred at the state level as a result of NPIs, assuming a linear scale. In order to make the results generalisable to larger population contexts, one would need to account for significant differences in NPI uptake and effectiveness across various populations and countries \cite{banholzer_estimating_2022}. Finally, the learned adaptive interventions do not guarantee global optimality, and solutions with even higher NHB are possible in principle.

Despite these limitations, the proposed approach can be effectively used to support policy- and decision-makers. On the one hand, the simulation of various pandemic scenarios, distinct WTP and compliance thresholds, as well as different demographic profiles,  can inform policy makers on the cost-effectiveness and possible trade-offs achieved by adaptive interventions, trained by the reinforcement learning method coupled with a calibrated ABM. On the other hand, the adaptive interventions may be compared against actual real-world data, so that the detected discrepancies may identify the divergence of some underlying assumptions, further elucidating the required responses and necessary adjustments. This may be particularly relevant in case of new variants of concern emerging during the anticipated endemic phase of the COVID-19. For example, the framework is directly applicable to modelling pandemic responses to the Omicron variant. Such an application would require straightforward changes in the ABM parameters, e.g., transmissibility, etc. (see Appendix: Table \ref{tab:calibrated_parameter_values})), vaccination efficacies, and the clinical burden rates (recovery distribution and fatality rate, see Appendix \ref{sec:SM_ABM}: \nameref{sm_recovery_and_fatality}). This is a subject of future work. Importantly, the proposed framework can enable a comprehensive evaluation of the role played by two key thresholds (WTP and $SD_{max}$), offering insights into the interplay between individual human behaviour and the emergent social dynamics during pandemics.

\section{Methods}
We propose a framework to search for cost-effective SD interventions balancing the health effects (averted DALY losses) and the costs associated with SD intervention measures. Decisions on SD interventions are taken with respect to a maximal SD level, determining a population fraction complying with the stay-at-home restrictions aimed to control the COVID-19 transmission in a typical Australian town. Our framework comprises three main components: (i) a method to evaluate the cost-effectiveness of SD intervention measures, (ii) an agent-based model (ABM) to simulate the effect of these interventions on the progression of the COVID-19 disease, and (iii) an RL algorithm to optimise an adaptive SD intervention simulated within the ABM and evaluated in terms of cost-effectiveness. The following sections describe these components in further detail. Our study did not involve experiments on humans/human data or the use of human tissue samples. The anonymised census data, which is related to the build of the ABM, are publicly available from the Australian Bureau of Statistics.

\subsection{Net Health Benefit}
\label{sec:nhb}

In order to evaluate the cost-effectiveness of NPI interventions, we quantify the net health benefit (NHB)~\cite{net_health_benefit, heitjan_fiellers_2000}. The NHB captures the difference between the health effect of a new intervention and the comparative health effect, given the associated cost incurred at some pre-defined cost-effectiveness thresholds. The cost and the health effect of the new intervention are measured against the ``null'' intervention, that is, in presence of some baseline interventions which are not subject to evaluation~\cite{WHO_undertaking_2003}. In our study, the null set comprises only the base interventions, i.e., case isolation (CI), home quarantine (HQ), and travel (border control) restrictions (TR). Hence, we evaluate cost-effectiveness of the  NPIs shaped by social distancing (SD), beyond that of the CI, HQ and TR interventions. The rate modulating the health effects' comparison is called ``willingness to pay'' (WTP), defined as a maximum monetary threshold that the society accepts as the cost of an additional unit of health gained due to the new intervention. The NHB of the SD intervention is defined as follows:
\begin{ceqn}
\begin{equation}
    \label{eq:nhb_true_mean}
    \text{NHB} = \mu_{E_{SD}} - \frac{\mu_{C_{SD}}}{\lambda}
\end{equation}
\end{ceqn}
where $\mu_{E_{SD}}$ is the mean of the health effect $E_{SD}$ produced by the SD intervention, $\mu_{C_{SD}}$ is the mean of the cost $C_{SD}$ incurred by this intervention, and $\lambda$ is the WTP set by policy makers or public health programs.

The corresponding health effect $E_{SD}$ of the SD intervention is computed by comparing the health losses averted by the evaluated intervention to the losses of the null intervention: 
\begin{ceqn}
\begin{equation}
    \label{eq:health_loss}
    E_{SD} = L_{0} - L_{SD} 
\end{equation}
\end{ceqn}
\noindent where $L_{0}$ and $L_{SD}$ are the health losses for the null and SD interventions respectively (see Fig. \ref{fig:health_effect} for illustration). In this study, we quantify health losses using Disability-Adjusted Life Year (DALY) approach recommended by the World Health Organization (WHO)~\cite{WHO_estimating_2003, neumann_systematic_2016}. Specifically, the years of life lost due to premature mortality (YLL) are combined with the years of life lived with disability (YLD), producing a single quantity expressing the burden of disease in time units: 
\begin{ceqn}
\begin{equation}
    \text{DALY} = \text{YLL} + \text{YLD}
\end{equation}
\end{ceqn}

For each infected individual (represented by an agent in the ABM), YLL is calculated as the difference between the life expectancy and the year of death if this agent dies as a result of the COVID-19. The second term, YLD, is measured by the duration of the disease within an infected agent who recovers from the COVID-19 (adjusted by a disability weight representing the disease severity). For non-infected agents, $\text{YLL} = 0$ and $\text{YLD} = 0$, under the assumption that the COVID-19 has not affected their health conditions. In this study, we also assumed that a life year lost due to the COVID-19-related death and an impacted year lived with disease for non-fatal cases are equally important (that is, we set the disability weight equal to 1). In addition, no age weighting and discounting on future health benefits~\cite{WHO_estimating_2003} were applied in our calculation for DALY, following \cite{wyper_burden_2021} and \cite{singh_disability-adjusted_2022}. The health impacts were calculated at the population level, accumulating the single measurements from all agents in our ABM.

Furthermore, the cost of the evaluated intervention is estimated under the assumption of the equal distribution of the total cost across the agents. When an SD intervention is imposed over a population fraction defined by some SD compliance level, the corresponding cost is assumed to be proportional to this fraction. For example, an intervention with the SD level of 50\% is assumed to cost half as much as the full lockdown at the SD level of 100\%. A scaling in proportion to the number of impacted individuals is applied in approximating the weekly intervention costs for a typical town, given the intervention costs  estimated at \$1.4 billion per week for the entire Australian economy (i.e., the entire population)~\cite{higginson_covid-19:_2020}. 

The NHB approach allows us to comparatively evaluate the cost-effectiveness of various interventions which may significantly differ in their costs and corresponding health effects. Consequently, it enables to derive adaptive SD interventions by gradually changing the SD levels in a direction that increases the cost-effectiveness. Thus, the NHB can be easily used by a reinforcement learning process exploring the search-space for more cost-effective interventions.

\subsection{Willingness to pay}
\label{sec:WTP}

Prior studies considered a broad range of the WTP levels. For example, the cost per quality-adjusted life year (QALY) can be estimated as the probability that the respondent will reject the bid values \cite{turnbull_empirical_1976, mcdougall_understanding_2020}. The estimates by this study resulted in: JPY 5.0 million in Japan (US\$41,000 per QALY), KNW 608 million in the Republic of Korea (US\$74,000 per QALY), NT 2.1 million in Taiwan (US\$77,000 per QALY), £23,000 in the UK (US\$36,000 per QALY), AU\$64,000 in Australia (US\$47,000 per QALY), and US\$62,000 per QALY in the USA.

Another approach determined that, on average, the cost per DALY averted was related to the Gross Domestic Product (GDP) per capita. For instance, the cost was 0.34 times the GDP per capita in low Human Development Index (HDI) countries, 0.67 times the GDP per capita in medium HDI countries, 1.22 times the GDP per capita in high HDI countries, and 1.46 times the GDP per capita in very high HDI countries~\cite{daroudi_cost_2021}. For Australia, this would correspond to the cost in the range of AU\$93,197.9 = US\$67,735.6 (or, 1.22 x  US\$55521.0) and AU\$111,531.9 = US\$81,060.7 (or, 1.46 x US\$55521.0). These estimates are produced using data from World Bank \cite{world_bank_gdp_capita} and International Monetary Fund \cite{imf_exchange} for the average 5-year GDP per capita and USD-AUD exchange rate, respectively.

Another accepted approach is to represent the WTP threshold as the (consumption) value that a society attaches to a QALY~\cite{bobinac_valuing_2013}. This societal perspective was followed by the contingent valuation approach which valued QALYs under uncertainty for the Dutch population, producing the range from €52,000 to €83,000 (approximately, AU\$82,409.9 - AU\$131,538.8).

A recent study in the Australian context used a range of WTP up to US\$300,000 (or AU\$412,771.9) per health-adjusted life year (HALY).  It specified preferable COVID-19 intervention policies in three ranges: (i) up to US\$20,000 (AU\$27,518.1), (ii) from US\$30,000 (AU\$41,277.2) to US\$240,000 (AU\$330,217.4), and (iii) above US\$240,000~\cite{Blakely_2020}.

These studies informed the choice of the WTP thresholds used in our analysis. In particular, we considered three WTP thresholds: \$10K per DALY, \$50K per DALY and \$100K per DALY.

\subsection{Agent-Based Model for COVID-19 Transmission and Control}
\label{sec:ABM}

In order to simulate transmission and control of the COVID-19 pandemic in Australia we used a well-established ABM~\cite{amtrac_19,amtrac-code-zenodo-2021,chang2022simulating}, calibrated to the Delta variant (B.1.617.2), and modified to capture a fluctuating adherence to social distancing as well as more refined vaccination coverage. The original ABM included a large number of individual agents representing the entire population of Australia in terms of demographic attributes, such as age, residence and place of work or education. In re-calibrating and validating this model, we used a surrogate population of New South Wales (7,485,860 agents), while the primary simulations, coupled with the RL algorithm, employed a surrogate population of 2,393 agents representing the population of a small Australian local area (e.g., a town), generated to match key characteristics of the Australian census carried out in 2016. The ABM is described in detail in Appendix \ref{sec:SM_ABM}: \nameref{sec:SM_ABM}, and here we only summarise its main features.

Each agent belongs to a number of mixing contexts (household, community, workplace, school, etc.) and follows commuting patterns between the areas of residence and work/education. The commuting patterns are obtained from the Australian census and other datasets available from the Australian Bureau of Statistics (ABS)~\cite{zachreson2018urbanization,fair2019creating,zachreson2020interfering}.
The transmission of the disease is simulated in discrete time steps, with two steps per day: daytime for work/education interactions, and nighttime for all residential and community interactions. The contact and transmission probabilities vary across contexts and ages.

The disease progression within an agent is simulated over several disease-related states, including Susceptible, Infectious (Asymptomatic or Symptomatic), and Removed. All agents are initialised as Susceptible. When an agent is Infectious, other susceptible agents sharing some mixing context with the agent may become infected, and infectious after some latent period. An age-dependent fraction of agents progresses through the disease asymptomatically.  The transmission probabilities are determined at each step, given the agents' mixing contexts, as well as their symptomaticity. The probability of transmission from an Infectious agent varies during the time since the exposure, growing to a peak of infectivity and then declining during the agent's recovery. At the end of the infectious period, the agents change their state to Removed (i.e., recovered or diseased), which excludes the agent from the Susceptible population. Thus, re-infections are not simulated, given that the simulated timeframe is relatively short (19 weeks following the first week during which the social distancing intervention is triggered, as mentioned below).

A pandemic scenario is simulated by infecting some agents. During calibration and validation, these agents are selected (``seeded'') in proximity to an international airport~\cite{amtrac_19,chang2022simulating}. During the primary simulations of each outbreak in a small Australian town, we seeded all initial cases within this area, according to a binomial sampling process, described in Appendix \ref{sec:SM_ABM}: \nameref{sec:SM_scenarios_seeding}. The seeding process is terminated when cumulative cases exceed a predefined threshold, simulating an imposition of travel restrictions around the town. At this point, the infections may continue to spread only as a result of the local transmission.

A vaccination rollout scheme is implemented in two modes: (i) a progressive rollout mode (i.e., reactive vaccination) used to validate the model with the actual data from the Sydney outbreak during June-November 2021, and (ii) a pre-emptive mode used to simulate pandemic scenarios controlled by NPIs, assuming that some population immunity has been already developed in response to past vaccination campaigns. Both modes assume hybrid vaccinations with two vaccines: Oxford/AstraZeneca (ChAdOx1 nCoV-19) and Pfizer/BioNTech (BNT162b2), concurring with the Australian campaigns during 2021~\cite{mass_vaccination_and_lockdown,chang2022simulating}. 

Different NPIs are simulated: case isolation, home quarantine, and social distancing interventions~\cite{amtrac_19,chang2022simulating}. Case isolation and home quarantine are assumed to be the baseline interventions, activated from the simulation onset. Social distancing (i.e., ``stay-at-home'' restrictions) is only triggered when cumulative cases surpass a specific threshold. Unlike previous implementations of the ABM, the compliance of agents, bounded by a given SD level, is simulated heterogeneously and dynamically, with Bernoulli sampling used to determine  whether an agent is compliant with the SD intervention at any given simulation step (within the total limit on the fraction of compliant agents). Vaccination states and compliance with NPIs modify the transmission probabilities in the corresponding mixing contexts, thus affecting spread of the outbreak. 

Importantly, the health effects resulting from the COVID-19 pandemic are captured by aggregating the high-resolution data simulated at the agent level. Unlike other studies which estimate the outcomes only at the end, we quantify the health losses after every simulation day, by measuring probable age-dependent deaths and total probable  impacted days for newly infected agents. This temporal resolution allows us to construct a decision process evaluating social distancing interventions in a way compatible with the RL method. Specifically, each decision point includes a state (i.e., information describing the current pandemic situation across all agents), an action (e.g., a decision setting a level of compliance with social distancing below the limit $SD_{max}$), and the associated outcome.

\vspace*{-1mm}
\subsection{Reinforcement Learning-based Search for Cost-effective NPIs}
\vspace*{-1mm}

Our framework for optimising the cost-effectiveness of SD interventions includes two typical RL components: a decision-maker and an environment, as shown in Fig.~\ref{fig:learning_framework}.
The decision-maker is configured as a neural network~\cite{sutton_reinforcement_2018, sym11020290,R_DDPG,structure_preserving} that can make decisions on the SD compliance levels (within the limit $SD_{max}$), given the decision-maker's observation of the environment. The environment comprises the ABM which simulates effects of these decisions on the transmission and control of the COVID-19 within a typical Australian town, as described in previous section.
Our objective is to learn the decision-making neural network based on the interactions between the decision-maker and the environment, so that cost-effectiveness of the SD intervention is maximised.

In our setting, once the outbreak starts, the decisions are assumed to be made every week, concurring with the time resolution adopted in other studies~\cite{cost_effectiveness_policy_iteration,india_lockdown,gb_school_closure}. At a decision point $t$, the decision-maker takes a (partial) observation of the environment (denoted by $o_t$), and selects its action $a_t$ aiming to cost-effectively control the ongoing outbreak. An observation characterises the current pandemic state, including the detected incidence (asymptomatic and symptomatic), prevalence, and the count of recoveries and fatalities.
Once decision $a_t$ is made setting the SD compliance for the next week, the ABM environment simulates the SD intervention associated with $a_t$ and its effects during the period from the decision point $t$ to the next decision point $t+1$. This simulation determines the economic costs incurred during the period (i.e., one week) and the associated health losses (averted DALY). These quantities constitute the reward signal, providing feedback to the decision-maker. At the next decision point this feedback is used to evaluate the choice of $a_t$.  

The interactions between the decision-maker and the environment start when the number of cumulative cases exceeds a threshold triggering the SD interventions, and continues until the end of the simulation period (e.g., includes $N = 19$ decision points $t \in \{0, 1, 2, ..., N\}$).
The interactions form a sequence of observations, actions, and rewards, registered at multiple decision points. The RL algorithm samples from this sequence, perform its optimisation, and updates the decision neural network. In general, this sampling step can be carried out at every decision point as the new data are collected, or be delayed depending on the algorithm.  

All interactions between the decision-maker and the environment form an ``episode'' in the learning process of optimal SD interventions. The learning process  involves multiple episodes independently following each other. The total reward generated during an episode, or the episodic reward, is expected to grow as the learning process continues, evidencing that the learned NPIs generate higher NHB. The learning process continues until only minimal improvements in the episodic reward are observed, marking convergence of the RL algorithm. The resultant decision neural network can then be used unchanged during the evaluation phase. The results and analysis presented in section \ref{sec:results} are based on the evaluation phase.

\begin{figure}[h]
    \centering
    \includegraphics[scale=0.4]{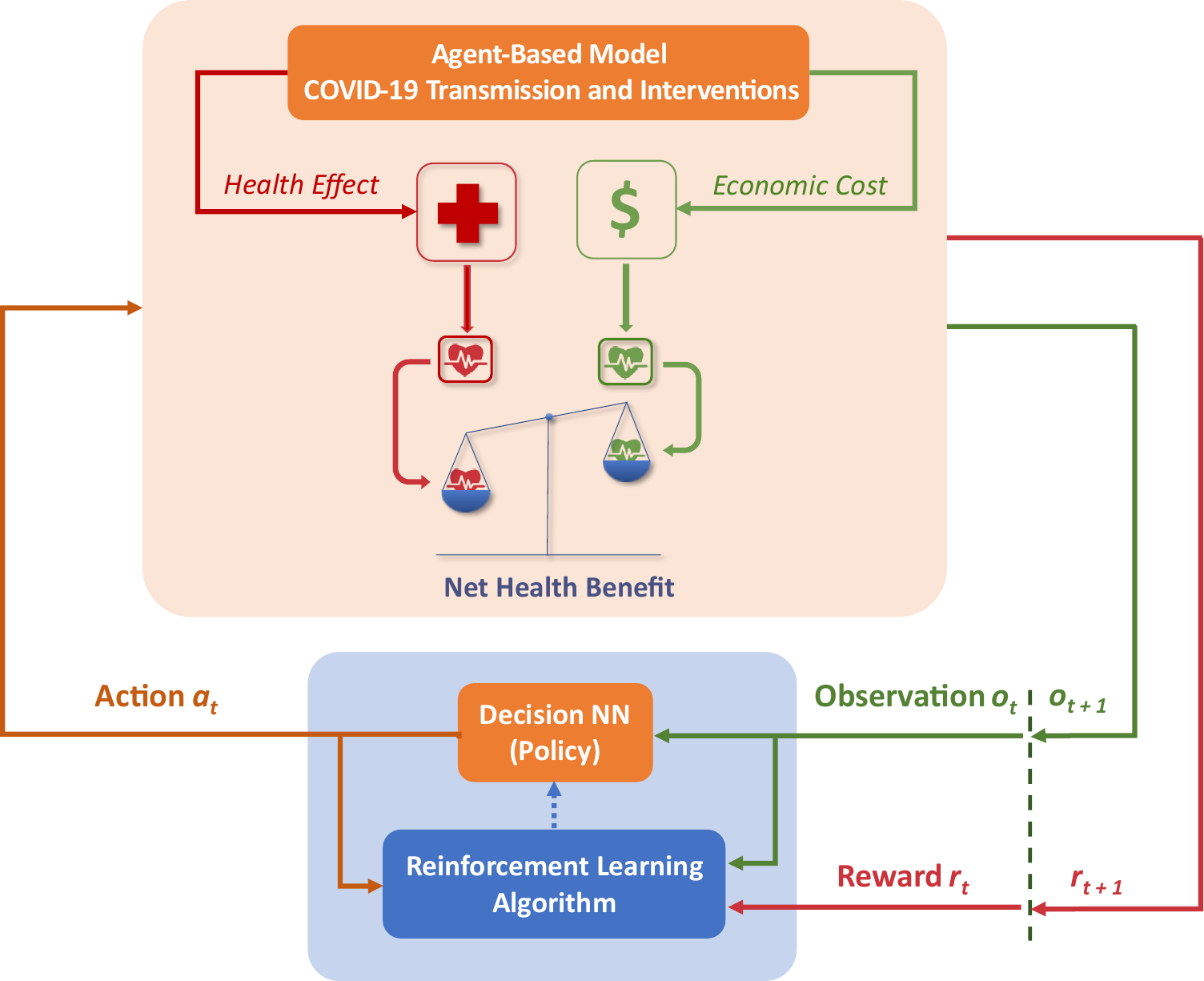}
    \caption{RL-based framework for optimising  cost-effectiveness of SD interventions.}
    \label{fig:learning_framework}
\end{figure}

The process of decision-making follows a Markov Decision Process (MDP) represented by a tuple $\langle S, O , P, A, R \rangle$. The set $S$ contains possible states of the environment. The decision-maker is assumed to  observe these states only partially, and the set $O$ contains all partial observations. The set $A$ contains all possible actions $a_t$ available to the decision-maker at each time step $t$. Each action $a_t \in A$ determines the corresponding SD compliance level $f(a_t)$ for the  SD intervention, imposed over the population between the time step $t$ and the next time step $t+1$. Formally, $f:A \rightarrow [0,1]$, e.g., $f(a^0_t) = 0$ for zero SD intervention $a^0$ at any time $t$. Note that $a_t$ defines the SD intervention applied in addition to baseline interventions, such as CI, HQ and TR which are always enabled by default. 

In general, the decision-maker can determine its action according to a stochastic policy $\pi:O \times A \rightarrow [0,1]$
, or a deterministic policy $\pi: O \rightarrow A$. 
In this study, we configure the decision-maker to follow a stochastic policy described by probability distribution $\pi(o,a)$. Given observation $o_t$ obtained at time step $t$, the action $a_t$ can be sampled from the policy distribution, denoted as $a_t \sim \pi(o_t,\cdot)$.  

Unlike other studies~\cite{sci_report_2020, india_lockdown}, which discretise the range of social distancing percentages, we defined $f(a_t)$ to be continuous in the interval $[0, SD_{max}]$, for some limit $0 \leq SD_{max} \leq 1$. The execution of an action $a_t$ at the state $s_t \in S$ constrains the 
environment dynamics developing 
between the time steps $t$ and $t+1$. The state transition probability, denoted by $P(s' | s,a): S \times A \times S \rightarrow [0,1]$, quantifies the chance of transition from the current state $s$ to the next state $s'$, following the execution of the action $a$. Thus, the probability $P(s' | s,a)$ reflects the pandemic dynamics controlled by the  interventions. After the action $a_t$ is executed, the environment produces the reward signal $r_{t+1} \in R$ so that the agent can reinforce its policy at the next time step $t+1$. Each reward $r_{t+1}$ is given by the corresponding health effects attained during the simulated period.

In order to optimise the SD interventions by maximising their NHB estimates over the entire simulation period of $N$ weeks, we use a period-wise approach maximising the following objective function (see Appendix \ref{SM:RL}: \nameref{sec:SM_obj_function} for further details):
\begin{ceqn}
\begin{equation}
    \label{eq:transformed_period_objective}
		\max_{\pi_{\theta}}  \ \	\mathop{\mathbb{E}}_{\substack{a_t \sim \pi_{\theta}(o_t, \cdot) \\ (s_t, a_t, s_{t+1}) \sim \tau \\ (s^0_t, a^0, s^0_{t+1}) \sim \tau^0}} \ \ \sum_{t=0}^{N} \left[ L(s^0_t, a^0) - L(s_t, a_t) - \frac{f(a_t) \ C^1}{\lambda} \right] \ ,
\end{equation}
\end{ceqn}
\noindent where $\pi_{\theta}$ is the policy shaped by parameters $\theta$; $L(s,a)$ is the health losses, measured in DALYs, resulting when the intervention at the SD level $a$ is applied to the environment at state  $s$; action $a_t$ is sampled from policy $\pi_{\theta}(o_t, \cdot)$ based on the environmental observation $o_t$; the transition from $s_t$ to $s_{t+1}$ belongs to the trajectory $\tau$ controlled by SD interventions $a_t$ (i.e., is sampled from a distribution of trajectories); the transition from $s^0_t$ to $s^0_{t+1}$  belongs to the uncontrolled trajectory $\tau^0$ shaped by null action $a^0$;  and $C^1$ is the mean cost for the full 100\% SD intervention between two consecutive time steps, with the full cost scaled down by the factor $f(a_t) \in [0, SD_{max}]$. The difference in the health losses between the trajectories $\tau^0$ and $\tau$, representing the health effects of the simulated SD intervention, is illustrated in Fig. \ref{fig:health_effect}.
\begin{figure}[h]
    \centering
    \includegraphics[scale=0.45]{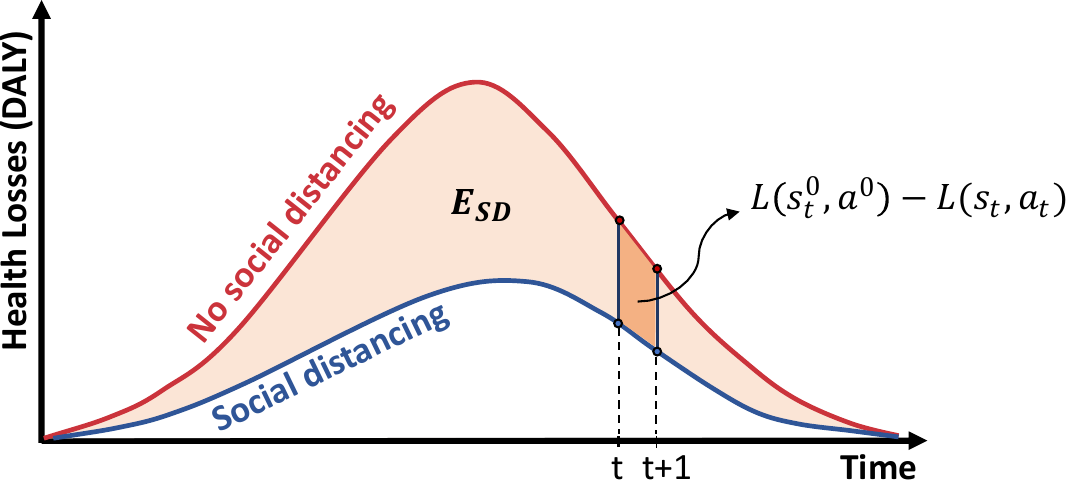}
    \caption{The health effect $E_{SD}$ is shown as the difference between health losses resulting from the null intervention without social distancing (red curve) and health losses averted by a social distancing intervention (blue curve). Period-wise health effect: $E_{SD} = \sum_{t=0}^N \left[ L(s^0_t,a^0) - L(s_t,a_t) \right]$.}
    \label{fig:health_effect}
\end{figure}

In order to maximise the objective function expressed by Eq. \ref{eq:transformed_period_objective}, we specify the reward signal for the action $a_t$ as follows:
\begin{ceqn}
\begin{equation}
    \label{eq:undiscounted_reward}
    r(s_t, a_t | s^0_t) = L(s^0_t, a^0) - L(s_t, a_t) - \frac{f(a_t) \ C^1}{\lambda}
\end{equation}
\end{ceqn}
Maximising the total received rewards along the trajectory $\tau$ is equivalent to maximising the objective expressed in Eq. \ref{eq:transformed_period_objective}, yielding the optimal decision-making policy $\pi^*$.

The ABM simulation is inherently stochastic, and hence, we use a discounted version of the accumulated rewards:
\begin{ceqn}
\begin{equation}
    \label{eq:discounted_objective_function}
   \max_{\pi_{\theta}} \ \ \mathop{\mathbb{E}}_{\substack{a_t \sim \pi_{\theta}(o_t, \cdot) \\ (s_t, a_t, s_{t+1}) \sim \tau \\ (s^0_t, a^0, s^0_{t+1}) \sim \tau^0}} \ \ \sum_{t=0}^{N}  \gamma^{t} r(s_t, a_t | s^0_t)   
\end{equation}
\end{ceqn}
\noindent where $\gamma \in (0,1)$ is the discount factor, and $r$ is the reward function defined by Eq. \ref{eq:undiscounted_reward}.

The policy $\pi_{\theta}$ is determined by a set of parameters $\theta$ which specify the weights of the decision neural network. A parameterised policy $\pi_{\theta}$ can be optimised by maximising a policy performance measure function $J(\theta)$. A canonical update for the parameter $\theta$ in each learning step $k$ follows the gradient ascent method~\cite{sutton_reinforcement_2018}, seeking to maximise the performance function $J(\theta)$:
\begin{ceqn}
\begin{equation}
    \label{eq:PG_gradien_ascent}
    \theta_{k+1} = \theta_{k} + \alpha \widehat{\nabla J(\theta_k)} 
\end{equation}
\end{ceqn}
\noindent where $\alpha$ is the learning rate for the update, and $\widehat{\nabla J(\theta_k)}$ is the estimation for the gradient of the performance function with respect to $\theta_k$. 

In our study, we used the Proximal Policy Optimisation (PPO) algorithm~\cite{schulman_proximal_2017} (see Appendix \ref{SM:RL}: \nameref{sec:SM_PPO}), aiming to avoid ``destructive large policy updates'' reported when the discounted objective function, defined by Eq. \ref{eq:discounted_objective_function}, is optimised directly~\cite{schulman_proximal_2017}. 
Specifically, we utilised the implementation of PPO for continuous actions provided by the Stable-Baselines3 library~\cite{stable-baselines3}. The convergence in the training for SD intervention policies, evidenced by improvement of the accumulated rewards over training episodes, is presented in Appendix \ref{SM:RL}: \nameref{sec:SM_Convergence}.

\section*{Acknowledgements}
This work was supported by the Australian Research Council grants DP220101688 and DP200103005 (MP). 
We would like to thank the developers of AMTraC-19~\cite{amtrac-code-zenodo-2021} for releasing the open source code of AMTraC-19 (i.e., the original ABM written in C++, which was extended and implemented in Python in our work). We also acknowledge the Sydney Informatics Hub for the use of the University of Sydney's high-performance computing cluster, Artemis.

\section*{Author contributions}
MP conceived and supervised the study. QN and MP developed the modelling framework. QN designed and implemented computational algorithms,  performed numerical simulations, and prepared the figures. MP and QN analysed the results, wrote the article, and approved the submitted version.

\begin{appendices}
\counterwithin{figure}{section}

\section{Agent-based Model}
\label{sec:SM_ABM}
We follow the agent-based modeling approach to simulate the transmission and control of COVID-19 on two scales: (i) a population of 7,485,860 agents to simulate the population of New South Wales, Australia, and (ii) a population of 2,393 agents to simulate the population of a local government area (LGA). The ABM simulates the interactions of agents within different social mixing contexts, with distinct contextual transmission probabilities calibrated to the Delta variant (B.1.617.2) of SARS-CoV-2. The surrogate population is generated based on the Australian Census and related data from the Australian Bureau of Statistics (ABS). The dynamics of the COVID-19 transmission incorporates both non-pharmaceutical interventions, e.g. social distancing, and vaccination rollout schemes. The original ABM, implemented in C++ programming language~\cite{amtrac_19} was extended with new features, including social distancing levels varying over time, two-dose vaccination, and estimation of the net health benefit in terms of the incurred economic costs and associated health effects, and implemented in Python programming language.


\subsection{Surrogate Population: Demographics}
Each agent in the surrogate population is individually simulated to represent an anonymous person with typical attributes, e.g., age, life expectancy, residence, workplace or school, according to the Australian Census, and other ABS datasets, as well as the Australian Curriculum and Assessment and Reporting Authority (ACARA) data. The age of an agent is sampled according to the 2016 ABS data on the estimated resident population by single year of age in New South Wales (NSW), Australia. The agents' life expectancy is assigned according to the 2016-2018 ABS data on life expectancy at birth by state and territory of usual residence. Agents' residences are artificially created based on the census of Statistical Local Areas (SLAs) and Collection Districts (CDs) defined by Australian Standard Geographical Classification (ASGC). Agents are assumed to follow recurrent travel patterns to usual destinations, e.g., workplaces or schools, and are expected to interact within communities and neighborhoods, as well as in their households. During each day, the agents interactions are split between two 12-hour routines: (i) contacts in the workplaces or the schools during daytime, and (ii) contacts in the communities, the neighborhoods, and the households during the nighttime. 

Both NSW and LGA populations are generated under the assumption of border closures with nearby states and LGAs respectively. We run the simulation with these surrogate populations to (a) validate the agent-based model with actual data in NSW, and (b) optimise social distancing interventions within a typical LGA, using a deep reinforcement-learning algorithm. 

\subsection{COVID-19 Transmission Model}
\subsubsection{Transmission Probability}
Transmission of SARS-COV-2 is assumed to be driven by interactions of the agents within their usual working, studying or living contexts. The model runs in discrete time steps (each is a 12-hour cycle) to represent the interactions (i) during the daytime cycle: workplaces (working groups) and schools (classes, grades, schools), and (ii) during the nighttime cycle: neighborhoods, communities, household clusters and households. When a susceptible agent is exposed, within a specific context, to potential infection spread by infectious agent(s), the transmission probability is determined by the infection probability of this context and the age of both susceptible and infectious agents, as described below, following prior studies~\cite{amtrac_19,chang2022simulating}. 

An agent can be in one of four states: Susceptible, Latent, Infectious (asymptomatic or symptomatic), and Removed (recovered or dead). The set $G_i$ contains all contexts to which agent $i$ belongs. Given a specific mixing context $g \in G_i$, the instantaneous probability $p^g_{j \rightarrow i}(n)$ is the probability that an infectious agent $j$ sharing the context $g$ with susceptible agent $i$ transmits the infection to agent $i$.  At the time step $n$, the infection probability for susceptble agent $i$ across the entire context $g$ is then calculated as follows:
\begin{ceqn}
\begin{equation}
    p^g_i(n) = 1 - \prod_{j \in A_g\backslash\{i\}} (1 - p^g_{j \rightarrow i}(n))
    \label{eq:infection_prob_within_a_group}
\end{equation}
\end{ceqn}
\noindent where $A_g\backslash\{i\}$ is the list of agents in the context $g$ excluding agent $i$. 
The instantaneous transmission probability is defined as follows:
\begin{ceqn}
\begin{equation}
\label{eq:individual_transmission_prob}
    p^g_{j \rightarrow i}(n) = \kappa \ f(n-n_j | j) \ q^g_{j \rightarrow i}    
\end{equation}
\end{ceqn}
\noindent where $\kappa$ is a global transmission scalar used to calibrate the reproductive number $R_0$, $n_j$ is the time step that agent $j$ becomes infected, and $f(n-n_j| j)$ is a function to characterise the infectivity of agent $j$ over time. For an uninfected agent $j$, $n-n_j < 0$ and $f(n-n_j | j)=0$. For an infected agent $j$, $n-n_j \geq 0$ and $f(n-n_j| j) \geq 0$. The natural history of the disease $f(n-n_j | j)$  is defined to follow a profile calibrated for B.1.617.2 (Fig.~\ref{fig:titer}), as described in prior work~\cite{chang2022simulating}. Asymptomatic agents are  modeled to be less infectious than symptomatic agents by a factor denoted by $\alpha_{\text{asymp}}$. Lastly, $q^g_{j \rightarrow i}$ is the daily probability of transmission from agent $j$ to agent $i$ at the agent $j$'s peak infectivity. The values of $q^g_{j \rightarrow i}$ in different mixing contexts are set in accordance to prior studies~\cite{amtrac_19,chang2022simulating} and specified in Table~\ref{tab:transmission_probability_at_peak_infectivity}.

\begin{table}[h]
    \centering
    \begin{tabular}{l|l|l}
    \textbf{Contact Group} &  \textbf{Type of Contact} & \textbf{Daily Transmission Probability}\\
    &  & \hspace{1.75cm} \textbf{($q^g_{j \rightarrow i}$)}\\
    \hline
    
    \rule{0pt}{3ex}Household (size 2)   & Any to child (0 - 18)             &  \hspace{1.7cm} 0.09335\\
                                        & Any to adult (19+)                &  \hspace{1.7cm} 0.02420\\
    \rule{0pt}{3ex}Household (size 3)   & Any to child (0 - 18)             &  \hspace{1.7cm} 0.05847\\
                                        & Any to adult (19+)                &  \hspace{1.7cm} 0.01495\\
    \rule{0pt}{3ex}Household (size 4)   & Any to child (0 - 18)             &  \hspace{1.7cm} 0.04176\\
                                        & Any to adult (19+)                &  \hspace{1.7cm} 0.01061\\
    \rule{0pt}{3ex}Household (size 5)   & Any to child (0 - 18)             &  \hspace{1.7cm} 0.03211\\
                                        & Any to adult (19+)                &  \hspace{1.7cm} 0.00813\\
    \rule{0pt}{3ex}Household (size 6)   & Any to child (0 - 18)             &  \hspace{1.7cm} 0.02588\\
                                        & Any to adult (19+)                &  \hspace{1.7cm} 0.00653\\[0.1cm]
    
    \hline                              
    \rule{0pt}{3ex}Household Cluster    & Child (0 - 18) to child (0 - 18)  &  \hspace{1.7cm} 0.00400\\
                                        & Child (0 - 18) to adult (19+)     &  \hspace{1.7cm} 0.00400\\
                                        & Adult (19+)  to child (0 - 18)    &  \hspace{1.7cm} 0.00400\\
                                        & Adult (19+)  to adult (19+)       &  \hspace{1.7cm} 0.00400\\[0.1cm]
    \hline
    \rule{0pt}{3ex}Working Group        & Adult (19+)  to adult (19+)       &  \hspace{1.7cm} 0.00400\\[0.1cm]
    
    \hline
    \rule{0pt}{3ex}School               & Child (0 - 18) to child (0 - 18)  &  \hspace{1.7cm} 0.00029\\
    Grade                               & Child (0 - 18) to child (0 - 18)  &  \hspace{1.7cm} 0.00158\\
    Class                               & Child (0 - 18) to child (0 - 18)  &  \hspace{1.7cm} 0.00865\\[0.1cm]
                           
    \hline
    \rule{0pt}{3ex}Neighborhood         & Any to child (0 - 4)          &  \hspace{1.7cm} $0.035 \times 10^{-5}$\\
                                        & Any to child (5 - 18)         &  \hspace{1.7cm} $1.044 \times 10^{-5}$\\
                                        & Any to adult (19 - 64)        &  \hspace{1.7cm} $2.784 \times 10^{-5}$\\
                                        & Any to adult (65+)            &  \hspace{1.7cm} $5.568 \times 10^{-5}$\\[0.1cm]
    
    \hline
    \rule{0pt}{3ex}Community            & Any to child (0 - 4)          &  \hspace{1.7cm} $0.872 \times 10^{-6}$\\
                                        & Any to child (5 - 18)         &  \hspace{1.7cm} $2.608 \times 10^{-6}$\\
                                        & Any to adult (19 - 64)        &  \hspace{1.7cm} $6.960 \times 10^{-6}$\\
                                        & Any to adult (65+)            &  \hspace{1.7cm} $13.92 \times 10^{-6}$\\[0.1cm]
    
    \hline
    \end{tabular}
    \caption{Daily transmission probability $q^g_{j \rightarrow i}$ from infected agent $j$ to susceptible agent $i$ in different contact groups. Numbers in brackets define age groups to categorise children or adults.}
    \label{tab:transmission_probability_at_peak_infectivity}
\end{table}

\begin{figure}
    \centering
    \includegraphics[scale=0.6]{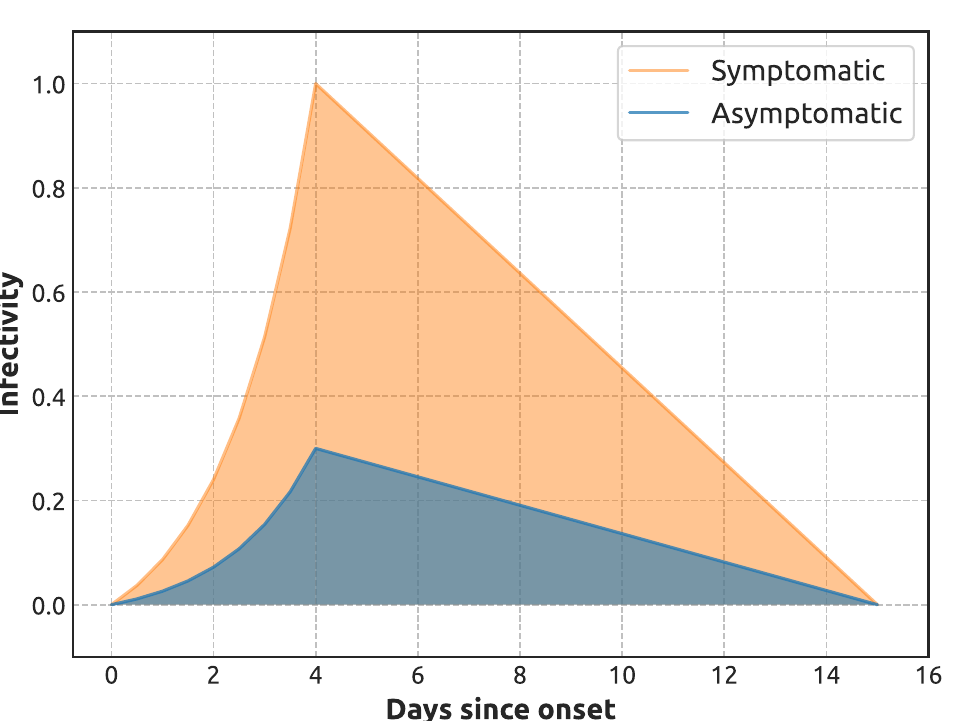}
    \caption{Simulated natural history of disease (calibrated to B.1.617.2). The agent infectivity is assumed to be exponentially increasing until its peak ($f(n-n_j | j) = 1.0$). In the post-peak period, the infectivity level linearly decreases to 0.0, the moment marking transition to Removed state.}
    \label{fig:titer}
\end{figure}

Finally, we derive the infection probability for agent $i$ at the time step $n$ across all shared contexts $G_i$:
\begin{ceqn}
\begin{equation}
\begin{split}
    p_i(n)  &= 1 - \prod_{g \in G_i(n)} \left( 1 - p^g_i(n) \right) \\
            &= 1 - \prod_{g \in G_i(n)} \prod_{j \in A_g\backslash\{i\}} \left( 1 - p^g_{j \rightarrow i}(n) \right)
\end{split}
\label{eq:general_infection_prob}
\end{equation}
\end{ceqn}
\noindent 
Bernoulli sampling, based on $p_i(n)$, is then used to determine whether a susceptible agent $i$ becomes infected at the end of each time step.

Since an infection can eventually lead to a symptomatic or asymptomatic case, we adjust the infection probability quantified by Eq.~\ref{eq:general_infection_prob} by  an additional factor $\sigma$ representing the fraction of symptomatic cases over the total cases:
\begin{ceqn}
\begin{equation}
    p_i^d(n) = \sigma(i) \ p_i(n)
    \label{eq:infection_prob_symptomatic}
\end{equation}
\end{ceqn}
\noindent where $\sigma(i)$ is a piecewise function dependent on the age of agent $i$. Specifically, $\sigma(i)$ is defined by a simple assignment of two values specified for adults ($\text{age} \geq 18$), $\sigma_a = 0.67$, and children ($\text{age} < 18$), $\sigma_c = 0.268$, see~\cite{chang2022simulating}.

\subsubsection{Recovery and Fatality}
\label{sm_recovery_and_fatality}
We use a truncated age-specific infection fatality rate (IFR) estimation~\cite{fatality_rate_original}, with a scaling adjustment for B.1.617.2~\cite{fatality_rate_delta}, to determine the probability of death for an infected agent:
\begin{ceqn}
\begin{equation}
    \label{eq:fatality_rate}
    \text{IFR}(\textit{age}) = \min \left(0.1, 0.0232 \times 10^{-3.27+0.0524 \times \textit{age}} \right)
\end{equation}
\end{ceqn}
Given a new infection, Bernoulli sampling is used, according to IFR expressed by Eq. \ref{eq:fatality_rate}, to determine whether the agent will recover or die  at the end of the disease progression (in Removed state). The outcome is then used to calculate the health effect for this agent, measured as disability-adjusted life years, i.e. DALYs. The effect of a death is estimated by the difference between the agent's current age and its expected life expectancy. The health loss of a recovered case is the onset-to-recovery time, estimated by a gamma distribution with the mean of 24.7 days and the coefficient of variation of 0.35, following~\cite{verity_2020}.

\subsection{Non-Pharmaceutical Interventions}
We model several non-pharmaceutical interventions (NPIs), e.g., case isolation (CI), home quarantine (HQ), school closure (SC), and social distancing (SD). Each NPI is modelled in terms of its compliance level, defined as the fraction of the population complying with this NPI, as well as the adjusted strength of interactions between a compliant agent and other agents sharing their mixing groups. The infection probability $p_i(n)$ for compliant agents is adjusted to account for the NPIs effects as follows:
\begin{ceqn}
\begin{equation}
    p_i(n) = 1 - \prod_{g \in G_i(n)} \prod_{j \in A_g\backslash\{i\}} (1 - F_g(j) \ p^g_{j \rightarrow i}(n))
    \label{eq:infection_prob_compliant_agent}
\end{equation}
\end{ceqn}
\noindent where $F_g(j) \neq 1$ is the strength of the interaction between agent $j$ and other agents in the shared context $g$. For non-compliant agents $j$, the interaction strength is unchanged: $F_g(j) = 1$.  

When agent $j$ complies with multiple NPIs, the value of $F_g(j)$ is preferentially assigned to only one NPI in accordance with the following order: CI, HQ, SD, SC. For example, if an infected agent is compliant with both CI and SD, its associated interaction strength is set according to CI: $F_g(j) = F_g^{\text{CI}}(j)$ as case isolation takes precedence over other interventions. Table~\ref{tab:NPIs} summarises the values of $F_g(j)$ for different interventions, as well as the baseline compliance levels of these interventions. Given these compliance levels, at each time step, the compliant and non-compliant agents  are randomly selected for each NPI according to Bernoulli process. While the compliance levels for CI, HQ, and SC are fixed in the simulation, the compliance level for SD (in short, the SD level) is controlled by a decision-making policy.

\begin{table}[h]
    \centering
    \begin{tabular}{l|c|c|c|c|c}
                                            &\multicolumn{5}{|c}{\textbf{Intervention}}\\[0.1cm]
        \hline
        \rule{0pt}{3ex}\textbf{Mixing Group}&\textit{CI}    &\textit{HQ}    &\textit{SD}    &\textit{SC (parent)}   &\textit{SC (child)}\\[0.1cm]
        \hline
        \rule{0pt}{3ex}Household            &1.0            &2.0            &1.0            &1.0                    &1.0\\
        \rule{0pt}{3ex}Household Cluster    &0.25           &0.25           &0.25           &0.5                    &0.5\\         
        \rule{0pt}{3ex}Working Group        &0.25           &0.25           &0.1            &0                      &-\\
        \rule{0pt}{3ex}School/Grade/Class   &0.25           &0.25           &0.1            &-                      &0\\
        \rule{0pt}{3ex}Neighborhood (CD)    &0.25           &0.25           &0.25           &0.5                    &0.5\\
        \rule{0pt}{3ex}Community (SLA)      &0.25           &0.25           &0.25           &0.5                    &0.5\\[0.1cm]
        \hline
        \rule{0pt}{3ex}\textbf{Compliance Level}     &0.7       &0.5        &-              &0.25                   &1.0\\[0.1cm]
        \hline
    \end{tabular}
    \caption{Interaction strengths and compliance levels for the considered NPIs across different mixing contexts.}
    \label{tab:NPIs}
\end{table}

\subsection{Vaccination Strategy and Vaccine Efficacy}

\subsubsection{Vaccination Strategy}
\label{sec:SM_Vaccination}

We simulate two vaccination rollout strategies in order to (i) validate our model against the actual epidemic data in NSW during an outbreak of the Delta variant over June-November 2021, and (ii) optimise adaptive SD interventions against future outbreaks within a local government area. The first objective follows (i) progressive vaccination strategy, capturing the vaccination dynamics in NSW in 2021, while the second objective is modelled with (ii) pre-pandemic vaccination strategy.

\textbf{Progressive vaccination.}
Approximating the actual vaccination campaign in NSW, the progressive rollout is modelled as a hybrid approach with two types of vaccines: Oxford/AstraZeneca (ChAdOx1 nCoV-19) and Pfizer/BioNTech (BNT162b2). Our simulation closely matches the number of first and second doses administered daily in NSW. The actual numbers are extracted from multiple COVID-19 vaccine rollout reports published daily by the Department of Health, Australian Government from 01 July 2021 to 27 October 2021~\cite{AUS_covid-19_vaccine_rollout}, as shown in Fig.~\ref{fig:NSW_mass_vaccination}. We also assume an equal distribution of ChAdOx1 nCoV-19 and BNT162b2 for individuals aged 16 and over, following another vaccine safety report from the Therapeutic Goods Administration (TGA) of the Department of Health, Australian Government, which specified that 12 million BNT162b2 doses and 10.8 million ChAdOx1 nCoV-19 were administered by 12 September 2021~\cite{TGA_vaccine_report}. For individuals aged 12-15, BNT162b2 is assumed to be the only administered vaccine, again in concordance with the adopted practice. In addition, for each agent, our simulated rollout strategy uses the same vaccine for dose 1 and dose 2, also in agreement with the practice in 2021. Our age-stratified daily vaccine allocation strategy for different age-groups (12-15, 16-49, 50-69, and 70+) is designed to satisfy these constraints.

\begin{figure}[H]
    \centering
    \includegraphics[width=\textwidth]{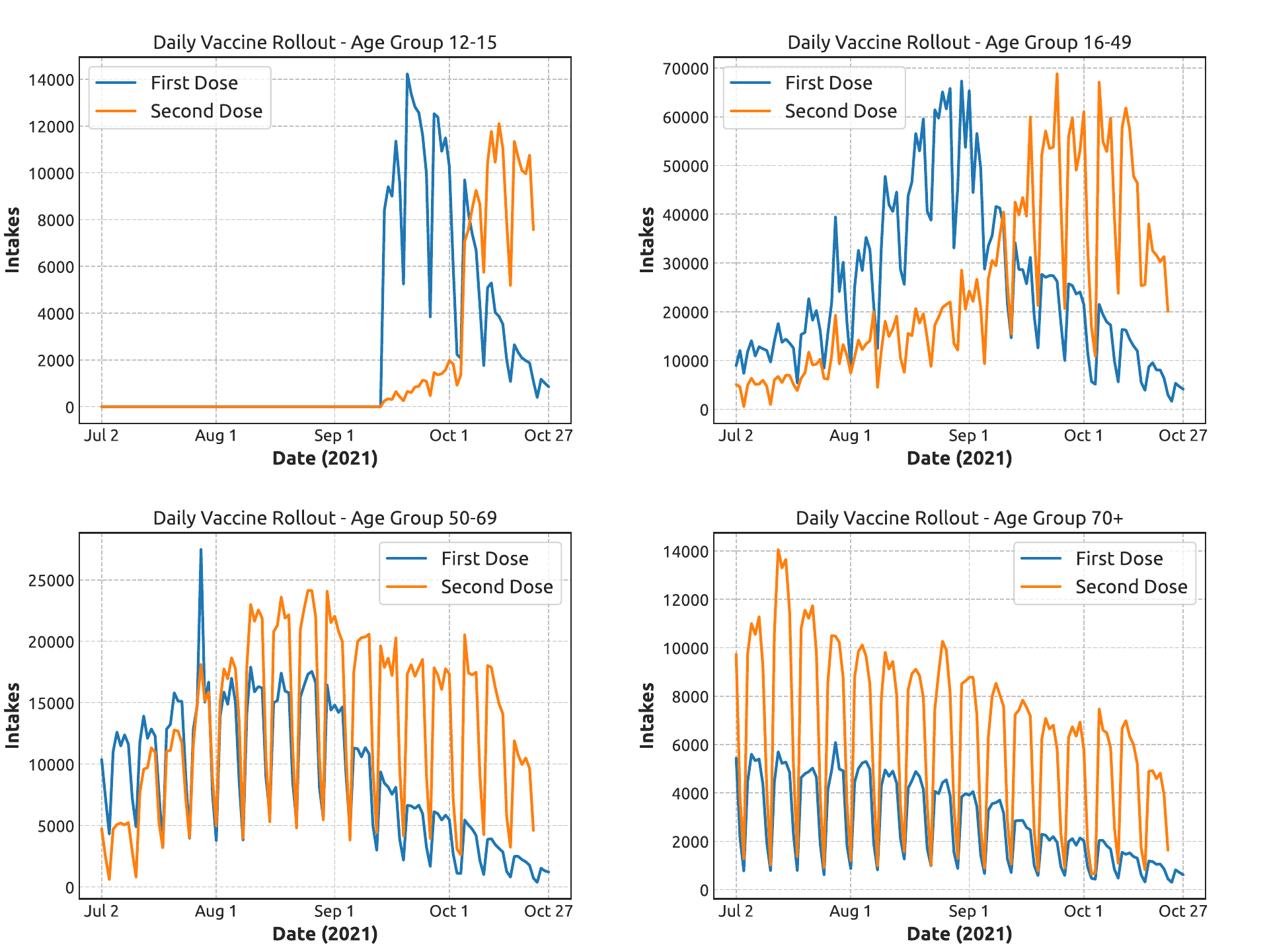}
    \caption{Number of first and second doses administered daily in New South Wales, Australia between 2 July 2021 and 27 October 2021. Source data are extracted from COVID-19 Vaccine Roll-out Reports published daily by the Department of Health, Australian Government during this period~\cite{AUS_covid-19_vaccine_rollout}.}
    \label{fig:NSW_mass_vaccination}
\end{figure}

\textbf{Pre-pandemic vaccination.}
Under this strategy, vaccine is rolled out before a pandemic, aiming to provide a significant portion of the population with some immunity. In our study, before the first simulation time step, two-dose vaccination is assigned to 85\% of adults and children. Pre-pandemic vaccination also assumes to follow a hybrid approach combining two vaccines: ChAdOx1 nCoV-19 and BNT162b2, with an approximately equal number of these two vaccines for individuals aged 16 and over, and predominantly BNT162b2 vaccine for individuals under 15 years of age.

\subsubsection{Vaccine Efficacy}
Following~\cite{mass_vaccination_and_lockdown, chang2022simulating}, our model is designed with different components of vaccine efficacy: efficacy against susceptibility (VEs), efficacy against susceptibility to disease (VEd), efficacy against infectiousness (VEi), as well as efficacy against death (VEp). Unlike~\cite{mass_vaccination_and_lockdown, chang2022simulating}, during the progressive vaccination, we vary these efficacy components over time, following a profile shown in Fig.~\ref{fig:VE_profile}. Vaccine efficacy is assumed to increase linearly from the days when the agent takes the doses (first, D1, or second, D2) and reach the maximum level of protection after a certain delay (Max\_D1 and Max\_D2 for first and second dose respectively). After taking the first dose, agents also need to wait a certain period of time (Min\_Delay) before registering for the next shot. Summary of our settings for these parameters is given in Table~\ref{tab:vaccine_profile_params} for ChAdOx1 nCoV-19 and BNT162b2. For the pre-pandemic vaccination strategy, we assume that all vaccinated agents get their both doses before an outbreak, and have sufficient time to build their full immunity against the COVID-19. The vaccine efficacy is assumed to be sustained at its highest level, once it is attained (which is realistic to assume for relatively short simulation horizons of approximately 20 weeks).

\begin{figure}[H]
    \centering
    \includegraphics[scale=0.65]{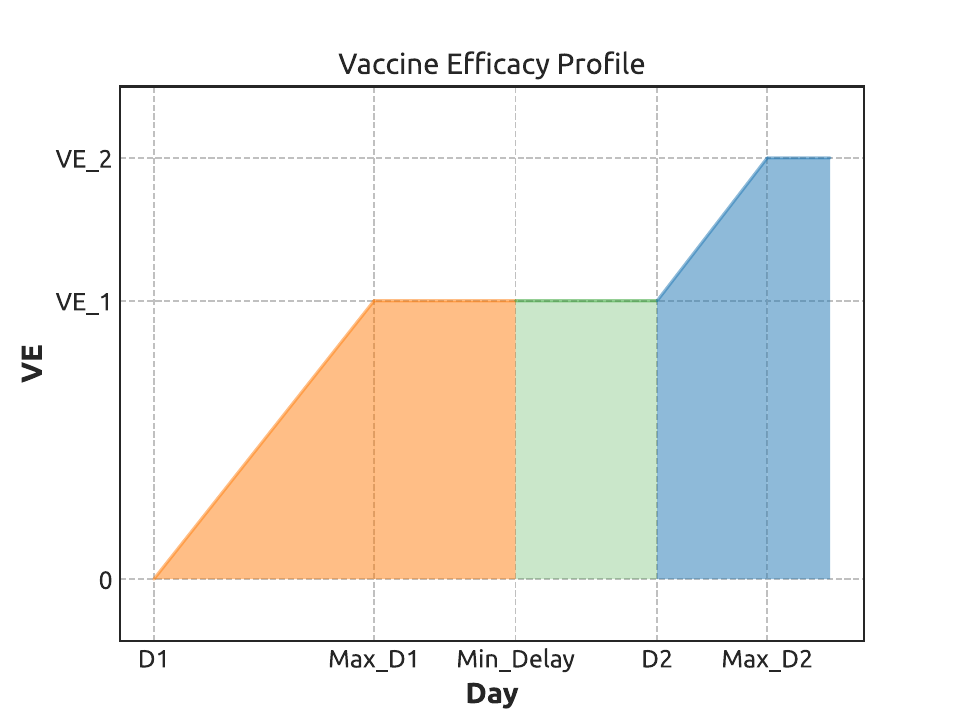}
    \caption{Vaccine Vaccine efficacy profile assumed for a two-dose vaccine rollout during the period between June 2021 and October 2021 in NSW, Australia. VE\_1 and VE\_2 denote the maximum  vaccine efficacy after dose 1 and dose 2, respectively. D1 and D2 denote the day of the first and second dose, while Max\_D1 and Max\_D2 denote the day when the maximum vaccine efficacy is reached for each dose. Min\_Delay denotes the delay between the doses.}
    \label{fig:VE_profile}
\end{figure}

\begin{table}[h]
    \centering
    \begin{tabular}{l|l|c|c}
        \rule{0pt}{3ex}VE Parameter &Description                                        &ChAdOx1 nCoV-19    &BNT162b2\\[0.1cm]
        \hline
        \rule{0pt}{3ex}D1           &Time when the first dose is taken                  &-                  &-\\ 
        Max\_D1                     &Time to build full immunity after the first dose   &D1 + 21 days       &D1 + 14 days\\
        Min\_Delay                  &Minimum time between the first and second doses    &28 days            &21 days\\
        D2                          &Time when the second dose is taken                 &-                  &-\\
        Max\_D2                     &Time to build full immunity after the second dose  &D2 + 14 days       &D2 + 7 days\\[0.1cm]
        \hline
    \end{tabular}
    \caption{Parameters setting the vaccine efficacy profile for ChAdOx1 nCoV-19 and BNT162b2.}
    \label{tab:vaccine_profile_params}
\end{table}

Given  vaccination components VEs and VEi, the infection probability defined by Eq.~\ref{eq:infection_prob_compliant_agent} is adjusted as follows:
\begin{ceqn}
\begin{equation}
    p_i(n) = \left( 1-\text{V}(\text{VEs}_{max}, n, n^{v}_{i}) \right) \left(1 - \prod_{g \in G_i(n)} \prod_{j \in A_g\backslash\{i\}} \left( 1 - (1-\text{V}(\text{VEi}_{max}, n, n^{v}_{j})) \ F_g(j) \ p^g_{j \rightarrow i}(n) \right) \right)
    \label{eq:general_trans_prob_vaccine_npi}
\end{equation}
\end{ceqn}
\noindent where $n^{v}_{i}$ and $n^{v}_{j}$ are the time steps when agent $i$ and agent $j$ take their latest vaccine shots, $\text{V}(\cdot)$ is a function returning vaccine efficacy values defined by the profile shown in Fig.~\ref{fig:VE_profile}, $\text{VEs}_{max}$ and $\text{VEi}_{max}$ are the maximum vaccine efficacies of the latest vaccine dose (dose 1 or dose 2) against susceptibility and infectiousness. For unvaccinated agents, Eq.~\ref{eq:general_trans_prob_vaccine_npi} simplifies with $V(\cdot)=0$.

The vaccine efficacy against disease (or illness), VEd,  affects the probability of generating a symptomatic case, denoted $p_i^d(n)$ for agent $i$. Eq. \ref{eq:infection_prob_symptomatic} is adjusted as follows:
\begin{ceqn}
\begin{equation}
    p_i^d(n) = \left( 1-V(\text{VEd}_{max}, n, n^{v}) \right)  \sigma(i) \ p_i(n)
    \label{eq:VEd}
\end{equation}
\end{ceqn}
\noindent where $\sigma(i)$ is a piece-wise function dependent on the age of agent $i$, setting the fraction of symptomatic agents.

Given the vaccination component VEp, the infection fatality rate of an infected agent at the time step $n$ is modified as follows:
\begin{ceqn}
\begin{equation}
    \label{eq:fatality_rate_vaccinated}
    \text{IFR}(\textit{age}, n) = \left( 1-V(\text{VEp}_{max}, n, n^{v}) \right) \ \min \left( 0.1, 0.0232 \times 10^{-3.27+0.0524 \ \textit{age}} \right)
\end{equation}
\end{ceqn}
\noindent where $\text{VEp}_{max}$ is the maximum vaccine efficacy against death attained after the 2nd dose.


Following~\cite{mass_vaccination_and_lockdown}, the values of the vaccine efficacy components are assigned to match the  clinical efficacy (VEc) against B.1.617.2~\cite{lopez_bernal_effectiveness_2021}, constrained as follows:
\begin{ceqn}
\begin{equation}
    \label{eq:VEc}
    \begin{split}
        \text{VEc} &= 1 - (1-\text{VEd})(1-\text{VEs})\\
            &= \text{VEs} + \text{VEd} - \text{VEs} \times \text{VEd}
    \end{split}
\end{equation}
\end{ceqn}
We solve Eq.~\ref{eq:VEc} for their central estimated values: $\text{VEs}=\text{VEd}=1-\sqrt{1-\text{VEc}}$. For BNT162b2, under assumption that $\text{VEc} = 0.7$ after dose 1, and $\text{VEc} = 0.9$ after dose 2, we set $\text{VEs} = \text{VEd} = 0.452$ after dose 1 and $\text{VEs} = \text{VEd} = 0.684$ after dose 2. For ChAdOx1 nCoV-19, under assumption that $\text{VEc} = 0.6$ after dose 1 and dose 2, we set $\text{VEs} = \text{VEd} = 0.368$. 

For the assignment of VEi, we assume the efficacy against onward transmission to be 0.45 after dose 1 and 0.5 after dose 2 for both types of vaccines~\cite{vaccine_effect_eng}. The vaccine efficacy against death is set as $\text{VEp}=0.71$ (after dose 1) and $\text{VEp}=0.92$ (after dose 2) for BNT162b2, and $\text{VEp}=0.69$ (after dose 1) and $\text{VEp}=0.90$ (after dose 2) for ChAdOx1 nCoV-19~\cite{scientific_advisory_group_for_emergencies}.

\subsection{Simulation Scenarios and Seeding Method}
\label{sec:SM_scenarios_seeding}

We simulate two different scenarios: (i) an outbreak in NSW developing alongside progressive vaccination, as well as NPIs including CI, HQ, SC and SD, and (ii) a potential outbreak in an SLA where 85\% of the population have been vaccinated (pre-pandemic vaccination). In scenario (ii), adaptive SD interventions are optimised by RL algorithm, while CI and HQ are always enabled as baseline measures. In both simulation scenarios, the outbreak starts once the first infections are `seeded'. 

For scenario (i), we seed the initial infections in the proximity of the Sydney International airport. Each simulated day, new infections are generated in the SLAs within a 50-kilometer radius of the Sydney's international airport, following a binomial distribution dependent on the average daily number of incoming passengers~\cite{chang2022simulating}. For scenario (ii), we seed initial infection within the considered SLA. 

The seeding continues until a threshold for cumulative cases, the travel restrictions (TR) threshold, is exceeded. During this seeding stage, only vaccination and baseline NPIs (CI, HQ) are simulated. The SD interventions  start only when another threshold of cumulative cases, the SD intervention threshold, is exceeded. For scenario (i), the TR threshold is set at 20, while the SD intervention threshold is set at 400 cases. For scenario (ii), both TR and SD intervention thresholds are set at 5.

\subsection{Model Calibration}
\label{sec:SM_calibration}

We calibrated the model to simulate the transmission of B.1.617.2 variant.  By varying ABM parameters, we explored a range of the reproduction number ($R_0$), aiming to attain $R_0$ at least twice as high as the reproduction number of the original SARS-CoV-2 variant~\cite{campbell_increased_2021}. For model validation based on NSW data, we aimed for $R_0$ to stay in the approximate range between 5.3 and 6.5. This range concurs with the  $R_0$ estimates for B.1.617.2 in the broad range, 3.2--8.0, reported previously~\cite{agency_for_clinical_innovation_living_2022, liu_reproductive_2021}, as well as the narrow range, 6.0--6.20, used in the Australian study with a similar ABM~\cite{chang2022simulating}. For optimisation of NPIs in an LGA, we aimed for $R_0$ to stay in a higher range between 7.0 and 8.0, in order to reflect a higher infectivity of variants causing future outbreaks.

$R_0$ is measured by the expected number of direct secondary cases from a typical infection in a entirely susceptible population \cite{diekmann_definition_1990}. In our ABM, in order to derive $R_0$, we randomly select an agent as the primary case, simulate transmissions, and count all direct secondary cases detected by the simulation. This process is repeated several hundred times. 

In order to eliminate the bias in selecting the primary case, we followed ``the attack rate pattern weighted index case'' method~\cite{germann2006mitigation,zachreson2020interfering} based on the age-specific attack rates [0.068, 0.173, 0.140, 0.461, 0.157] determined by the overall simulation for specific age groups: [0-4, 5-18, 19-29, 30-64, 65+]. The resultant value of $R_0$ for the validation scenario was 6.348 (95\% CI 5.858--6.839, $N=300$). 
For the optimisation scenario with higher infectivity, the resultant value of $R_0$ was 7.582 (95\% CI 7.457--7.706, $N=4,416$), as detailed in subsection~\ref{asymp_inf}.

Many studies have suggested that children are affected less severely by SARS-COV-2 than adults~\cite{liguoro_sars-cov-2_2020, davies_age-dependent_2020}. In our ABM, the probability of becoming a symptomatic case is age-dependent. The fraction of symptomatic cases in children was calibrated to be 60\% less than adults: $\sigma_{child} = 0.268$, while $\sigma_{adult} = 0.67$. The attack rate for children (under 18), simulated under $\sigma_{child} = 0.268$ without SD interventions, reached 0.24. This rate is in the range of [0.22, 0.29] reported by the National Centre for Immunisation Research and Surveilance (NCIRS) for Sydney outbreak in 2021 \cite{national_centre_for_immunisation_research_and_surveillance_covid-19_2021}.

The set of calibrated parameters is summarised in Table~\ref{tab:calibrated_parameter_values}.

\begin{table}[h]
    \centering
    \begin{tabular}{c|l|l}
        Parameters  &Description    &Value  \\
        \hline
        $\kappa$    &Global transmission scalar in Eq. \ref{eq:individual_transmission_prob}       &6.0\\
        \hline
        $\sigma$    &Probability that an infection becomes &0.67 (for adults)\\
                    &\hspace{0.5cm} a symptomatic case in Eq. \ref{eq:infection_prob_symptomatic} and \ref{eq:VEd} &0.268 (for children)\\
        \hline
        $\alpha_{\text{asymp}}$ &Factor for the reduction in transmissibity & 0.3 (validation)\\
        & \hspace{0.5cm} of an asymptomatic case as shown in Fig. \ref{fig:titer}& 0.5 (optimisation)\\
        \hline
        $T_{\text{lat}}$    & Latent period          & 0\\
        $T_{\text{inc}}$    & Incubation period      & 4 days\\
        $T_{\text{inf}}$    & Infectious period      & 15 days\\
        
        \hline
    \end{tabular}
    \caption{Calibrated ABM parameters.}
    \label{tab:calibrated_parameter_values}
\end{table}

\subsection{Sensitivity Analysis}

In order to check robustness of the ABM, we performed a local point-based sensitivity analysis. The analysis quantified the response of specific output variables to changes in key input parameters which were varied while keeping  other input variables at their default values. As described in Section \ref{sec:SM_calibration}, the default values were calibrated using the data of the third COVID-19 wave which started in NSW in June 2021. The following input parameters were investigated: SD intervention threshold, global transmission scalar ($\kappa$), infectious period ($T_{\text{inf}}$), the fraction of symptomatic cases in children ($\sigma_{child}$), and the reduction in infectivity of asymptomatic cases (i.e., asymptomatic infectivity $\alpha_{asymp}$). The peak incidence and the total fatalities were selected as the output variables. Figures~\ref{fig:sm_sa_threshold}--\ref{fig:sm_sa_asymptomatic_infectivity} traced responses of the output variables with respect to changes in the input parameters. 
This sensitivity analysis used the simulations for scenario (i), as described in Section~\ref{sec:SM_scenarios_seeding}. For each value of the input parameter, we independently simulated a fixed SD intervention specified at different levels: 30\%, 40\%, or 50\%, that is, $SD = 0.3$, $SD = 0.4$, or $SD = 0.4$. Other input parameters, unless varied themselves, are set at their default values determined by the calibration.

\subsubsection{SD Intervention Threshold}
The threshold of cumulative cases which triggers the SD interventions is an important input parameter shaping pandemic response. The outbreak in NSW started on 16 June 2021, following a long period without any confirmed locally acquired cases (more than a month since 5 May 2021~\cite{nsw_covid-19_cases}). The stay-at-home orders with various levels of restrictions  were progressively issued since late June 2021. Since the outbreak kept escalating, a tighter lockdown was announced in NSW on 9 July 2021~\cite{nsw_lockdown_order_july9}, when cumulative incidence reached 449 cases (detected between 16 June 2021 and 8 July 2021). Following this real-world account, we varied the SD intervention threshold in the range between 50 and 450 cases, with an increment step of 50. Figure~\ref{fig:sm_sa_threshold} traces the output variables, i.e. peak incidence and total fatalities, over the simulation period of 114 days.  For each simulated SD level (0.3, 0.4, 0.5), the peak incidence and the number of total fatalities are observed to grow with the increase in the SD intervention threshold from 50 to 450. 

For $SD = 0.3$, the peak incidence increases 4.06 times from the median value 866.5 (first quartile: 626.25, third quartile: 1240) to the median value 3514 (first quartile: 3305.25, third quartile: 3868.5). The total fatalities increase 4.36 times from the median value 496 (first quartile: 366.75, third quartile: 706.75) to the median value 2162.5 (first quartile: 1986, third quartile: 2284).
 
For $SD = 0.4$, the peak incidence increases 4.19 times from the median value 367.5 (first quartile: 344, third quartile: 550.5) to the median value 1540.5 (first quartile: 1380, third quartile: 1740). The total fatalities increase 4.5 times from the median value 218 (first quartile: 204, third quartile: 343.25) to the median value 981.5 (first quartile: 891.5, third quartile: 1098).

For $SD = 0.5$, the peak incidence increases 5.64 times from the median value 169 (first quartile: 141.75, third quartile: 204.5) to the median value 952.5 (first quartile: 831.25, third quartile: 1009.5). The total fatalities increase 6.83 times from the median value 89 (first quartile: 65.25, third quartile: 125) to the median value 608 (first quartile: 558.5, third quartile: 668.75).

In summary, a 9-fold increase in the SD intervention threshold linearly leads to an approximately 4 to 7 times increase in the output variables (peak incidence and total fatalities) across all simulations with fixed SD levels (30\%, 40\%, and 50\%). While this sensitivity is higher for lower value $SD = 0.3$, it remains moderate for higher considered SD levels, as shown in Fig.~\ref{fig:sm_sa_threshold}. Under $SD = 0.4$ or $SD = 0.5$, i.e., the compliance levels which have been retrospectively estimated for NSW at the time~\cite{chang2022simulating}, the observed sensitivity markedly diminishes for the SD intervention thresholds which exceed 300 cases and approac the threshold used in NSW (450 cases). This shows that the model is broadly robust to changes in the threshold, with the robustness strengthening in the policy-relevant range.

\begin{figure}
    \centering
    \includegraphics[width=\textwidth]{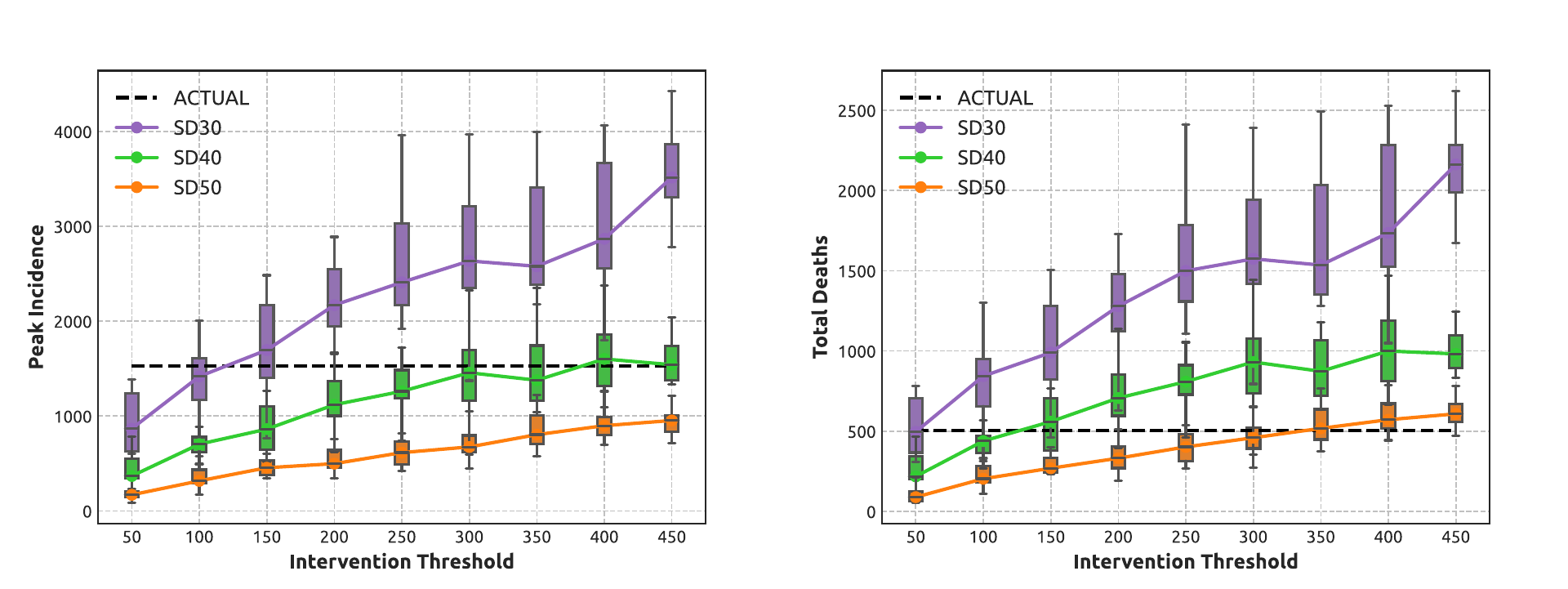}
    \caption{Local sensitivity analysis of the output variables (peak incidence and total number of fatalities) with respect to changes in the SD intervention threshold. The black dashed lines trace the actual peak incidence and the total number of fatalities in NSW during the simulated period. }
    \label{fig:sm_sa_threshold}
\end{figure}

\subsubsection{Global Transmission Scalar}

The global transmission scalar ($\kappa$, see Eq. \ref{eq:individual_transmission_prob}) was varied around its default value of 6.0, which resulted from the calibration process, as described in section \ref{sec:SM_calibration}. We varied $\kappa$ in the range between 5.6 and 6.4, with the increment step of 0.2. The simulated SD interventions (0.3, 0.4, 0.5) are all triggered at the threshold of 400 cumulative cases. Figure~\ref{fig:sm_sa_kappa} shows the corresponding changes in the peak incidence and total fatalities. 

For $SD = 0.3$, with the grow of $\kappa$, the peak incidence increases by  253.04\% relative to the peak incidence obtained at the lower bound $\kappa = 5.6$ (median 1663, first quartile: 1540.75, third quartile: 1819.5). Similarly,  the total fatalities increase by  227.63\% relative to the fatalities simulated at the lower bound $\kappa = 5.6$ (median 1046, first quartile: 922.5, third quartile: 1147). 

For $SD = 0.4$, the peak incidence linearly increases by 226.69\%, and the total fatalities linearly increase by 193.18\%. 

For $SD = 0.5$, the peak incidence almost linearly increases by 206.57\%, and the total fatalities almost linearly increase by 192.71\%. 

In summary, a 14.3\% growth in $\kappa$ from 5.6 to 6.4 increases the outputs approximately 2 to 2.5 times, indicating moderate to high sensitivity, as expected for the global transmission scalar which directly affects the reproduction number $R_0$. Nevertheless, the reported dependencies are linear within the policy-relevant range of $SD = 0.4$ to $SD = 0.5$, and the model is robust in the proximity to the default value $\kappa = 6.0$.

\begin{figure}
    \centering
    \includegraphics[width=\textwidth]{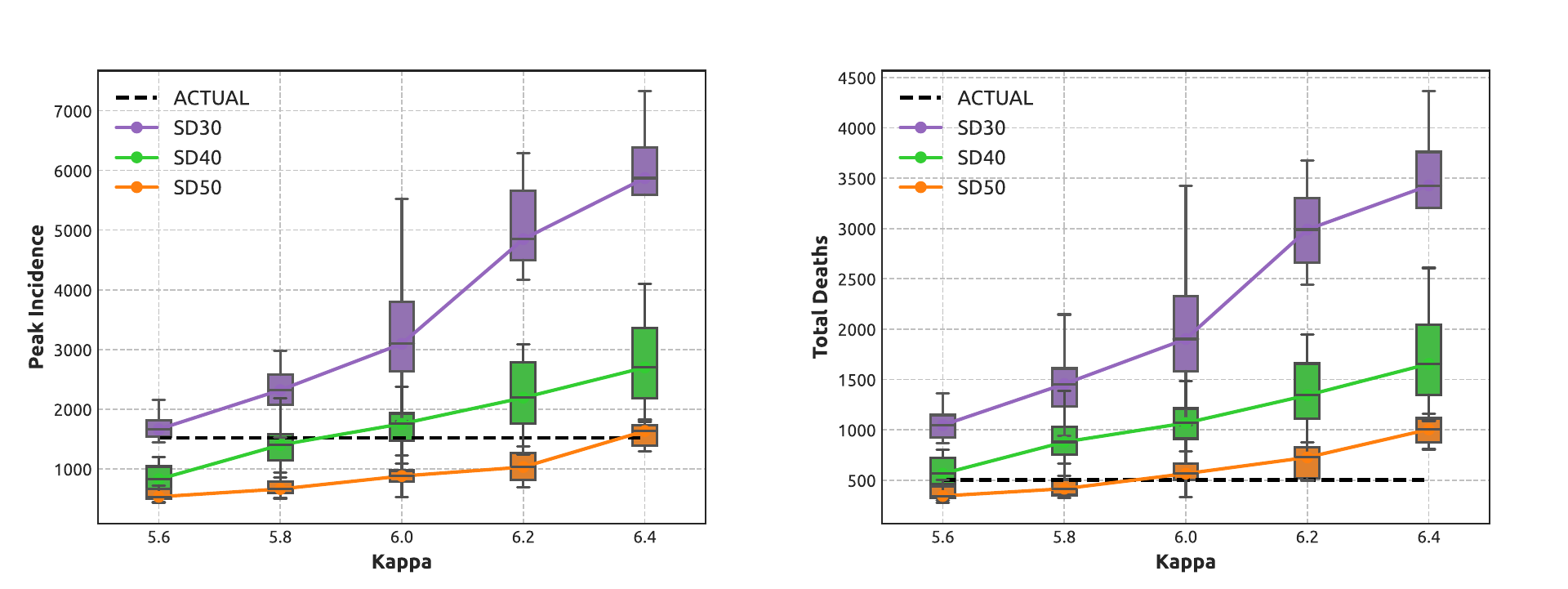}
    \caption{Local sensitivity analysis of the output variables (peak incidence and total number of fatalities) with respect to changes in the global transmission scalar ($\kappa$). The black dashed lines trace the actual peak incidence and the total number of fatalities in NSW during the simulated period. }
    \label{fig:sm_sa_kappa}
\end{figure}

\subsubsection{Duration of Infectious Period}

The infectious period $T_{\text{inf}}$ spans the entire natural history of the disease, with the infectivity rising to its peak and then linearly reducing to zero, see Fig.~\ref{fig:titer}. We explored the sensitivity of the two output variables with respect to changes in the duration for infectious period which was varied around the calibrated value (15 days). The comparison between the periods of 11, 13, 15 and 17 days was performed for $R_0 \approx 6.35$ (i.e., $\kappa = 6.0$), and the SD intervention threshold set at 400 cases, across three SD compliance levels (0.3, 0.4, 0.5). While varying the infectious period, we fixed the incubation period at 4 days, thus changing only the post-incubation period which was varied as 7, 9, 11 and 13 days. 

Figure~\ref{fig:sm_sa_infectious_period} shows a high sensitivity of the output variables (peak incidence and total fatalities) to the changes in the infectious period. For $SD = 0.3$, the peak incidence produced by $T_{\text{inf}} = 11$ days has the median value 452 (first quartile: 390.25, third quartile: 495.75), and increases dramatically by 1076.88\% at $T_{\text{inf}} = 17$ days. The number of total fatalities increases by 984.91\% (17 days) relative to fatalities at $T_{\text{inf}} = 11$  days (median: 275, first quartile: 215.25, third quartile: 327.5). Smaller but still significant sensitivities are also observed for $SD = 0.4$ and $SD = 0.5$.  These observations limit the model robustness to changes in the duration of infectious period within a narrow range around $T_{\text{inf}} = 15$ and within the policy-relevant range of $SD = 0.4$ to $SD = 0.5$. The observed sensitivity to $T_{\text{inf}}$ highlights the impact of the infectious period's duration on the pandemic scale, especially under modest social distancing levels.

\begin{figure}
    \centering
    \includegraphics[width=\textwidth]{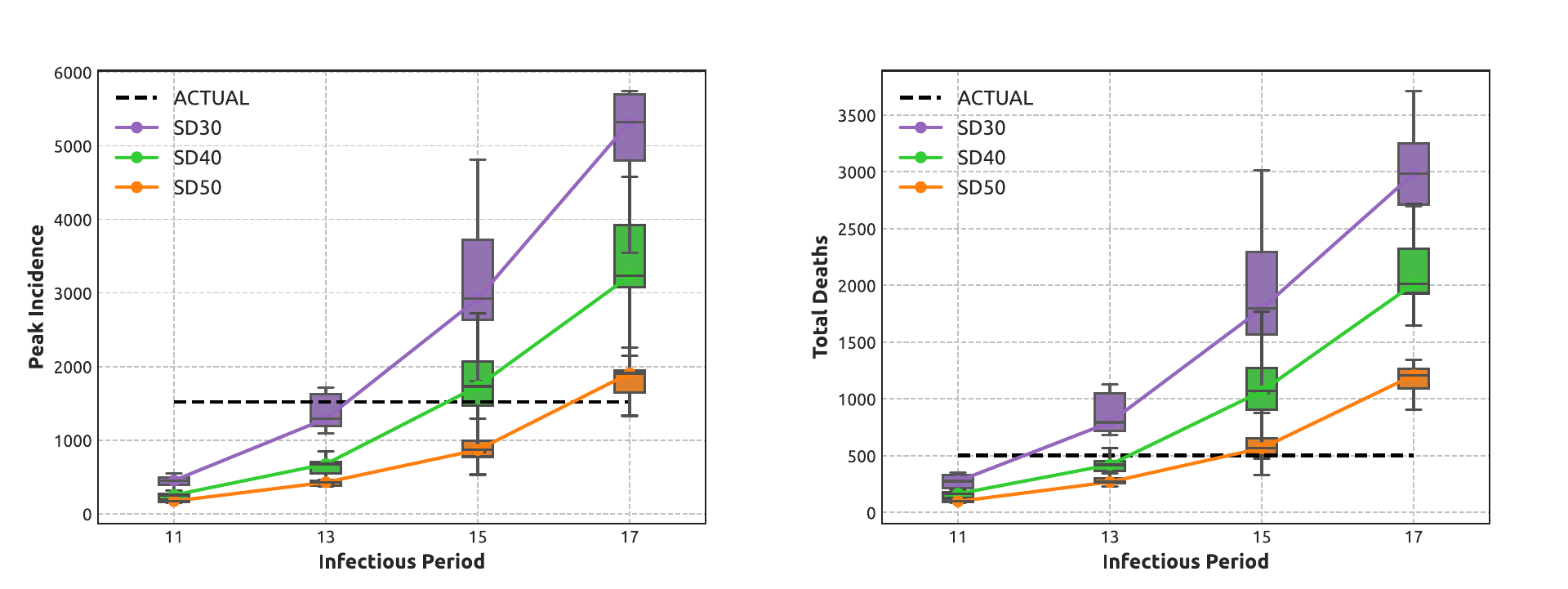}
    \caption{Local sensitivity analysis of the output variables (peak incidence and total number of fatalities) with respect to changes in the infectious period ($T_{\text{inf}}$). The black dashed lines trace the actual peak incidence and the total number of fatalities in NSW during the simulated period. }
    \label{fig:sm_sa_infectious_period}
\end{figure}

\subsubsection{Fraction of Symptomatic Cases in Children}

As the fraction of detected cases in children tends to increases with the B.1.617.2 variant compared to the original variant~\cite{macartney_transmission_2020, national_centre_for_immunisation_research_and_surveillance_covid-19_2021}, we investigated the sensitivity to $\sigma_{child}$. This input parameter was varied in the range [0.067, 0.268], with the increment step of 0.067. 

Figure~\ref{fig:sm_sa_child_asymp_fraction} shows that, as the fraction $\sigma_{child}$ increases, the peak incidence and the number of total fatalities do not exhibit much sensitivity for any SD compliance level, from $SD = 0.3$ to $SD = 0.5$. Across the entire range $\sigma_{child}$, i.e., [0.067, 0.268], the differences between the first and third quartiles of all boxplots (for a specific SD level) are relatively small, with the boxplots nearly overlapping. The low sensitivity of the outputs to changes in $\sigma_{child}$ confirms model robustness with respect to changes in the fraction of detected cases in children.

\begin{figure}
    \centering
    \includegraphics[width=\textwidth]{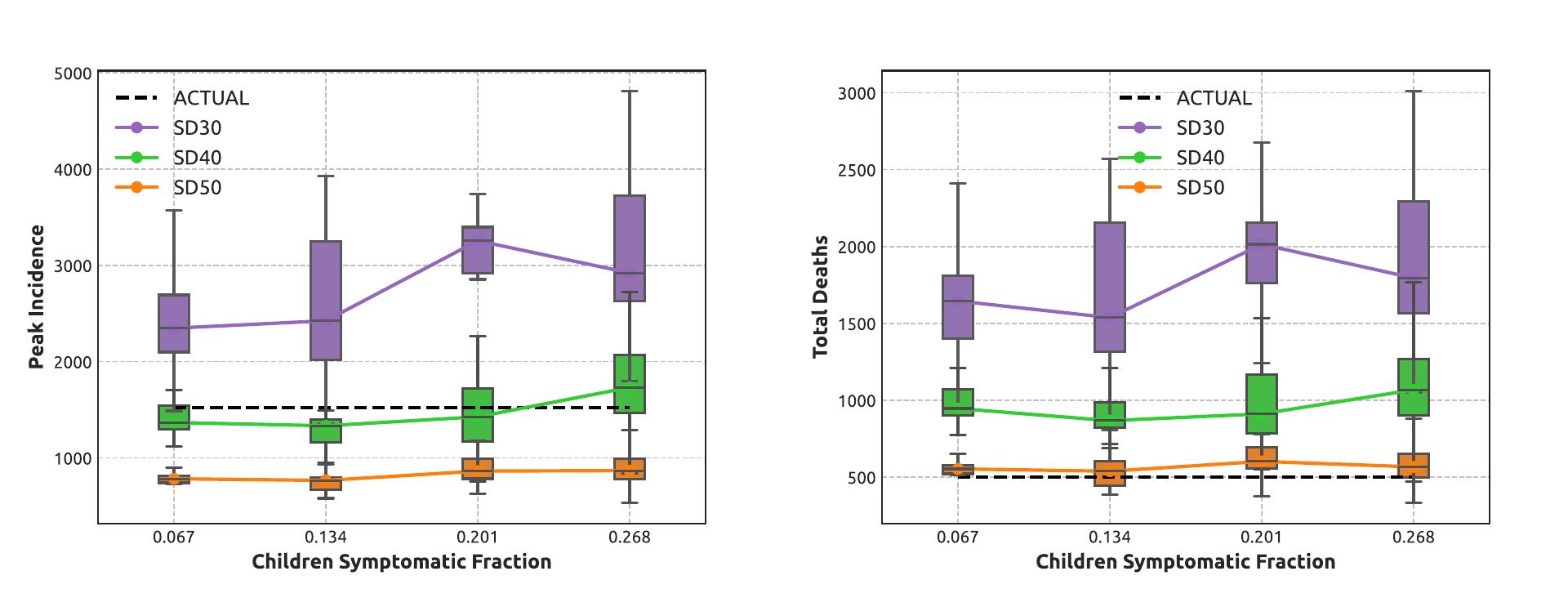}
    \caption{Local sensitivity analysis of the output variables (peak incidence and total number of fatalities) with respect to changes in the child symptomatic fraction ($\sigma_{child}$). The black dashed lines trace the actual peak incidence and the total number of fatalities in NSW during the simulated period. }
    \label{fig:sm_sa_child_asymp_fraction}
\end{figure}

\subsubsection{Asymptomatic Infectivity}
\label{asymp_inf}
In our model, asymptomatic cases are modelled as less infectious than symptomatic cases. The relevant parameter, asymptomatic infectivity $\alpha_{asymp}$, is the fraction specifying the infectivity of a typical infected asymptomatic case in comparison to the maximum level of infectivity in a typical symptomatic case (Fig.~\ref{fig:titer}). We varied $\alpha_{asymp}$ within the range [0.2, 0.5], with increment step 0.1. As shown in Fig.~\ref{fig:sm_sa_asymptomatic_infectivity}, the change in asymptomatic infectivity affects both output variables in all scenarios from $SD = 0.3$ to $SD = 0.5$. 

For $SD = 0.3$, the peak incidence at the lower bound $\alpha_{asymp} = 0.2$ has the median value 918.5 (first quartile: 841.5, third quartile: 1023.75), and increases by 1617\% when this parameter reaches the upper bound $\alpha_{asymp} = 0.5$. The number of total fatalities starts with the median value 556 (first quartile: 536.75, third quartile: 641) at the lower bound $\alpha_{asymp} = 0.2$, and increases by 1478.33\% at the upper bound $\alpha_{asymp} = 0.5$. 

For $SD = 0.4$ to $SD = 0.5$, however, the changes in output variables are less significant (about two and three times less, respectively). This points to the model robustness within the policy-relevant range of $SD = 0.4$ to $SD = 0.5$. 

Optimisation scenarios used an upper bound $\alpha_{asymp} = 0.5$, to reflect potentially higher asymptomatic infectivity. This setting corresponds to $R_0 = 7.582$, with 95\% CI 7.457--7.706, $N = 4,416$.

\begin{figure}
    \centering
    \includegraphics[width=\textwidth]{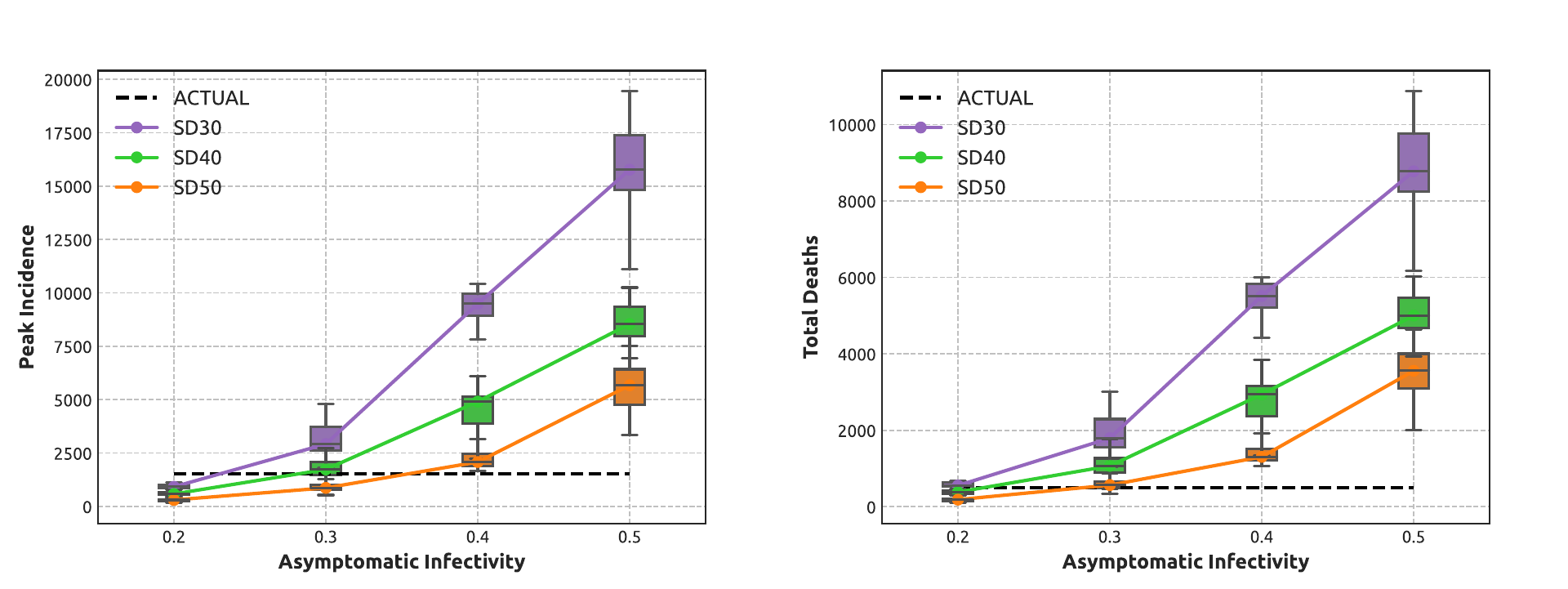}
    \caption{Local sensitivity analysis of the output variables (peak incidence and total number of fatalities) with respect to changes in the infectivity of asymptomatic cases ($\alpha_{asymp}$). The black dashed lines trace the actual peak incidence and the total number of fatalities in NSW during the simulated period. }
    \label{fig:sm_sa_asymptomatic_infectivity}
\end{figure}

\subsubsection{Summary}

The local sensitivity analysis showed robustness of the model to changes in several key input parameters: SD intervention threshold, global transmission scalar, duration of infectious period, fraction of symptomatic cases in children, and asymptomatic infectivity. The highest sensitivity of the outputs variables (the peak incidence and the number of total fatalities) was detected with respect to changes in the infectious period's duration and the asymptomatic infectivity. Nevertheless, the model was shown to be robust in the neighbourhood of the default parametrisation, especially within the policy-relevant range of the compliance with  stay-at-home orders, that is, $SD = 0.4$ to $SD = 0.5$.

\subsection{Model Validation}
\label{sec:SM_Validation}

The ABM was validated using the actual epidemic data in NSW during the period between 16 June 2021 and 27 October 2021~\cite{nsw_covid-19_cases,nsw_covid-19_deaths}. While the numbers of COVID-19-induced deaths and incidence are accessible for the entire period, the vaccination statistics  categorised into doses and age groups for NSW are only available for the period since 1 July 2021. As a result, in our simulation, we assume that the NSW vaccination coverage  was unchanged from 16 June 2021 to 1 July 2021 with the initial coverage mapped to 1 July 2021. The simulation of daily vaccinations starts on 2 July 2021 according to the profile matching Fig.~\ref{fig:NSW_mass_vaccination}. The other parameters for the COVID-19 transmission model are kept at their calibrated values, described in Section~\ref{sec:SM_calibration}.

The incidence and the daily fatalities are plotted in Fig.~\ref{fig:sm_validation1} and Fig.~\ref{fig:sm_validation2}, contrasting the actual time series and the  profiles simulated with different SD interventions ($SD = 0.3$ to $SD = 0.6$), each triggered by a threshold of 400 cumulative cases. As mentioned earlier, the range of stay-at-home compliance between $SD = 0.4$ and $SD = 0.6$ concurs with the retrospective analysis of the NSW outbreak~\cite{chang2022simulating} and is supported by the actual mobility reduction data ~\cite{covid19data_mobility}. Figures~\ref{fig:sm_validation1} and~\ref{fig:sm_validation2} show that the closest match to the actual incidence data is given by $SD = 0.4$ and $SD = 0.5$, while the new fatalities align best with $SD = 0.5$ and $SD = 0.6$. These results validate the ABM, enabling its use in the scenarios aimed to derive and optimise adaptive SD interventions.

\begin{figure}
    \centering
    \includegraphics[scale=0.65]{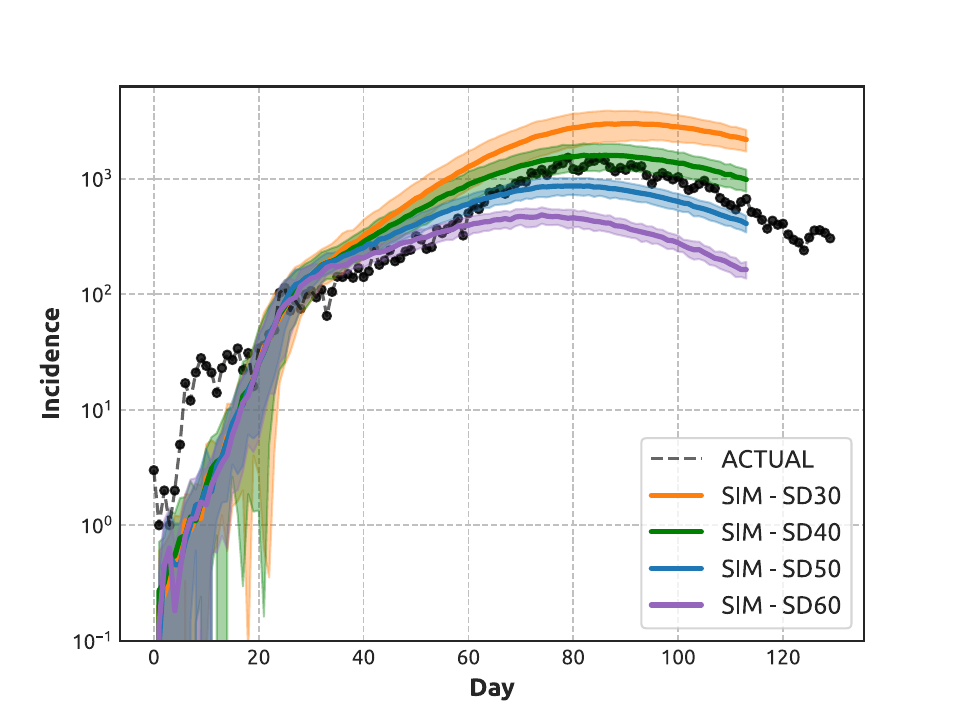}
    \caption{Model validation using actual COVID-19 incidence data in NSW, Australia, from 16 June 2021 (Day 0) to 12 October 2021 (Day 118), shown in log scale. The actual time series, shown in black, is obtained from NSW Health datasets~\cite{nsw_covid-19_cases}. The mean values and confidence intervals of simulations are shown in colour, varying across different levels of compliance with social distancing, from 30\% to 60\% (over 20 runs per scenario). }
    \label{fig:sm_validation1}
\end{figure}

\begin{figure}
    \centering
    \includegraphics[scale=0.65]{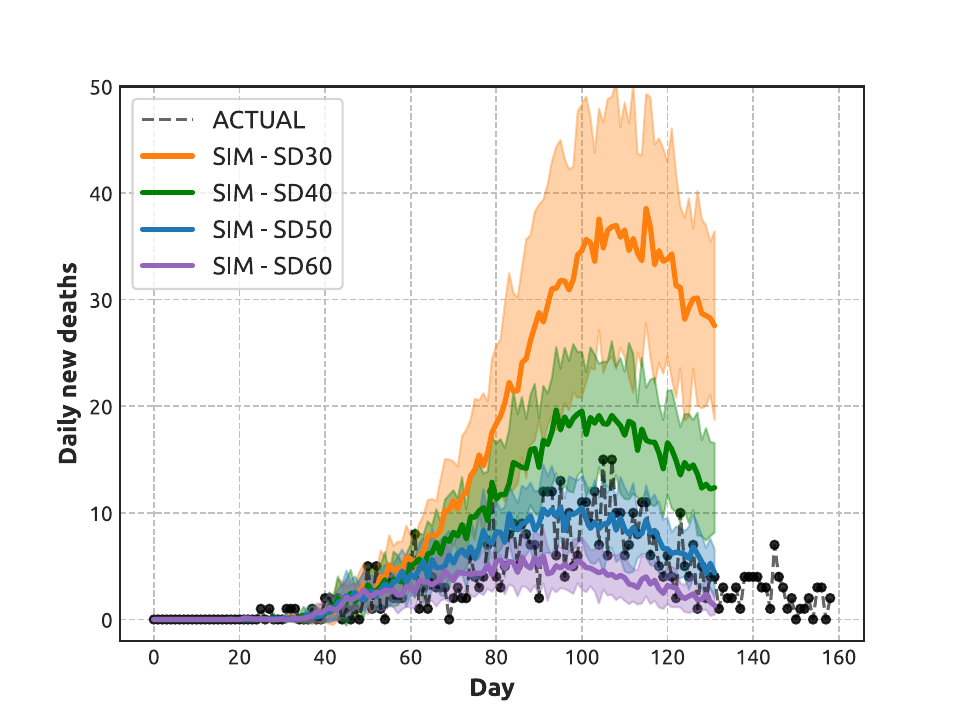}
    \caption{Model validation using actual COVID-19 data on daily fatalities in NSW, Australia, from 16 June 2021 (Day 0) to 12 October 2021 (Day 118). The actual time series, shown in black, is obtained from NSW Health datasets~\cite{nsw_covid-19_deaths}. The mean values and confidence intervals of simulations are shown in colour, varying across different levels of compliance with social distancing, from 30\% to 60\% (over 20 runs per scenario). }
    \label{fig:sm_validation2}
\end{figure}


\section{Reinforcement Learning Approach to Search for Cost-Effective Social Distancing Policies}
\label{SM:RL}

\subsection{Period-wide NHB Objective Function}
\label{sec:SM_obj_function}
This section provides details for the setup of the period-wide net health benefit (PW-NHB) objective function used in Section 4.3. In order to evaluate the cost-effectiveness of an SD intervention policy, the NHB is calculated for the entire simulation period:
\begin{ceqn}
\begin{equation}
    \begin{split}
    \label{eq:period_objective}
        \operatorname*{PW-NHB} &= \mu_{{E}_{SD}} - \frac{ \mu_{C_{SD}}}{\lambda} \\
        &= \mathop{\mathbb{E}}_{\substack{a_t \sim \pi_{\theta}(o_t, \cdot) \\ (s_t, a_t, s_{t+1}) \sim \tau \\ (s^0_t, a^0, s^0_{t+1}) \sim \tau^0}} \ \ \sum_{t=0}^{N}L(s^0_t, a^0) - \sum_{t=0}^{N}L(s_t, a_t) - \frac{\sum_{t=0}^{N}C(s_t, a_t) - \sum_{t=0}^{N}C(s^0_t, a^0)}{\lambda} \\
    \end{split}
\end{equation}
\end{ceqn}
\noindent where $\mu_{E_{SD}}$ and $\mu_{C_{SD}}$ are the mean values of measurements of the health effect and the cost of SD intervention during the entire simulation period; $\pi_{\theta}$ is the policy shaped by parameters $\theta$; action $a_t$ is sampled from policy $\pi_{\theta}(o_t, \cdot)$ based on the environmental observation $o_t$; the transition from $s_t$ to $s_{t+1}$ belongs to the epidemic trajectory $\tau$ controlled by SD interventions $a_t$; the transition from $s^0_t$ to $s^0_{t+1}$  belongs to the uncontrolled trajectory $\tau^0$ shaped by null action $a^0$; $L(s,a)$ is the health losses, measured in DALYs, resulting when the intervention at the SD level $a$ is applied to the environment at state $s$; $C(s, a)$ is the cost incurred between two consecutive time steps when an SD intervention $a$ is applied together with the base NPIs, i.e. CI and HQ, at the state $s$ of the environment; and $\lambda$ is the willingness-to-pay (WTP) parameter.

As mentioned in Section 4.1, the cost of SD intervention is estimated in proportion to the number of compliant agents:
\begin{ceqn}
\begin{equation}
\begin{split}
    \label{eq:period_economic_cost}
    \sum_{t=0}^{N}C(s_{t}, a_t) - \sum_{t=0}^{N}C(s^0_{t}, a^0) &\approx \sum_{t=0}^{N} \left[ C(s_{t}, a^0) + f(a_t) \ [C(s_{t}, a^1) - C(s_{t}, a^0)] - C(s^0_t, a^0) \right] \\
    &\approx \sum_{t=0}^{N} \left[ f(a_t) \ [C(s_{t}, a^1) - C(s_{t}, a^0)] \right] \\
    &\approx \sum_{t=0}^{N} f(a_t) \ C^1
\end{split}
\end{equation}
\end{ceqn}
\noindent where $f(a_t)$ is the SD compliance level associated with the action $a_t$, $a^0$ is the action for zero SD intervention, and $a^1$ is the action for full 100\% SD intervention. In general, the baseline costs for CI and HQ, i.e., $C(s_{t}, a^0)$ and $C(s^0_{t}, a^0)$, may vary according to the number of incident cases. However, in this study, for simplicity, we assumed their values to be constant at every time step. In addition, we also assumed that the mean value $C^1$ of the costs for the full 100\% SD intervention, i.e. $C(s_{t}, a^1) - C(s_{t}, a^0)$, is also known. These assumptions contribute to the approximations taken in Eq. \ref{eq:period_economic_cost}. Hence, the objective function expressed by Eq. \ref{eq:period_objective} is reduced as follows:
\begin{ceqn}
\begin{equation}
    \label{eq:sm_transformed_period_objective}
		\operatorname*{PW-NHB} \ \ \approx  \ \	\mathop{\mathbb{E}}_{\substack{a_t \sim \pi_{\theta}(o_t, \cdot) \\ (s_t, a_t, s_{t+1}) \sim \tau \\ (s^0_t, a^0, s^0_{t+1}) \sim \tau^0}} \ \ \sum_{t=0}^{N} \left[ L(s^0_t, a^0) - L(s_t, a_t) - \frac{f(a_t) \ C^1}{\lambda} \right]
\end{equation}
\end{ceqn}
\subsection{Empirical Convergence in the Training of SD Policies}
\label{sec:SM_Convergence}
Figures \ref{fig:rewards_training_SD30}-\ref{fig:rewards_training_SD70} show convergence of the training phase under different combination of WTPs (\$10,000/DALY, \$50,000/DALY, and \$100,000/DALY) and $SD_{max}$ (30\%, 50\%, and 70\%).

\begin{figure}[ht]
    \centering
    \includegraphics[width=\textwidth]{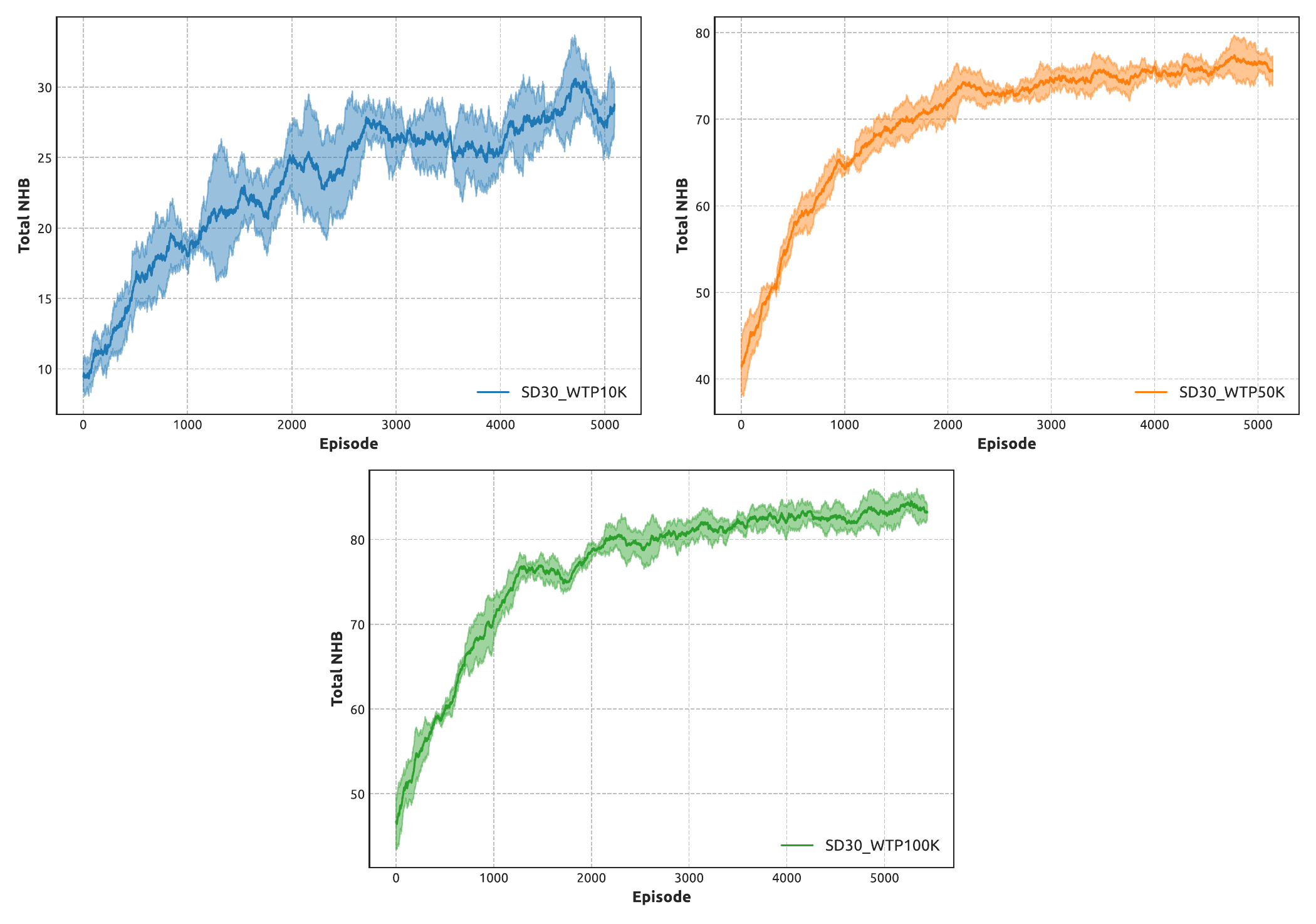}
    \caption{Accumulated episodic rewards generated over training episodes by the SD interventions trained under different combinations of WTP and $SD_{max}=0.3$. Solid curves represent the mean values and the shaded areas represent standard deviation.}
    \label{fig:rewards_training_SD30}
\end{figure}

\begin{figure}[ht]
    \centering
    \includegraphics[width=\textwidth]{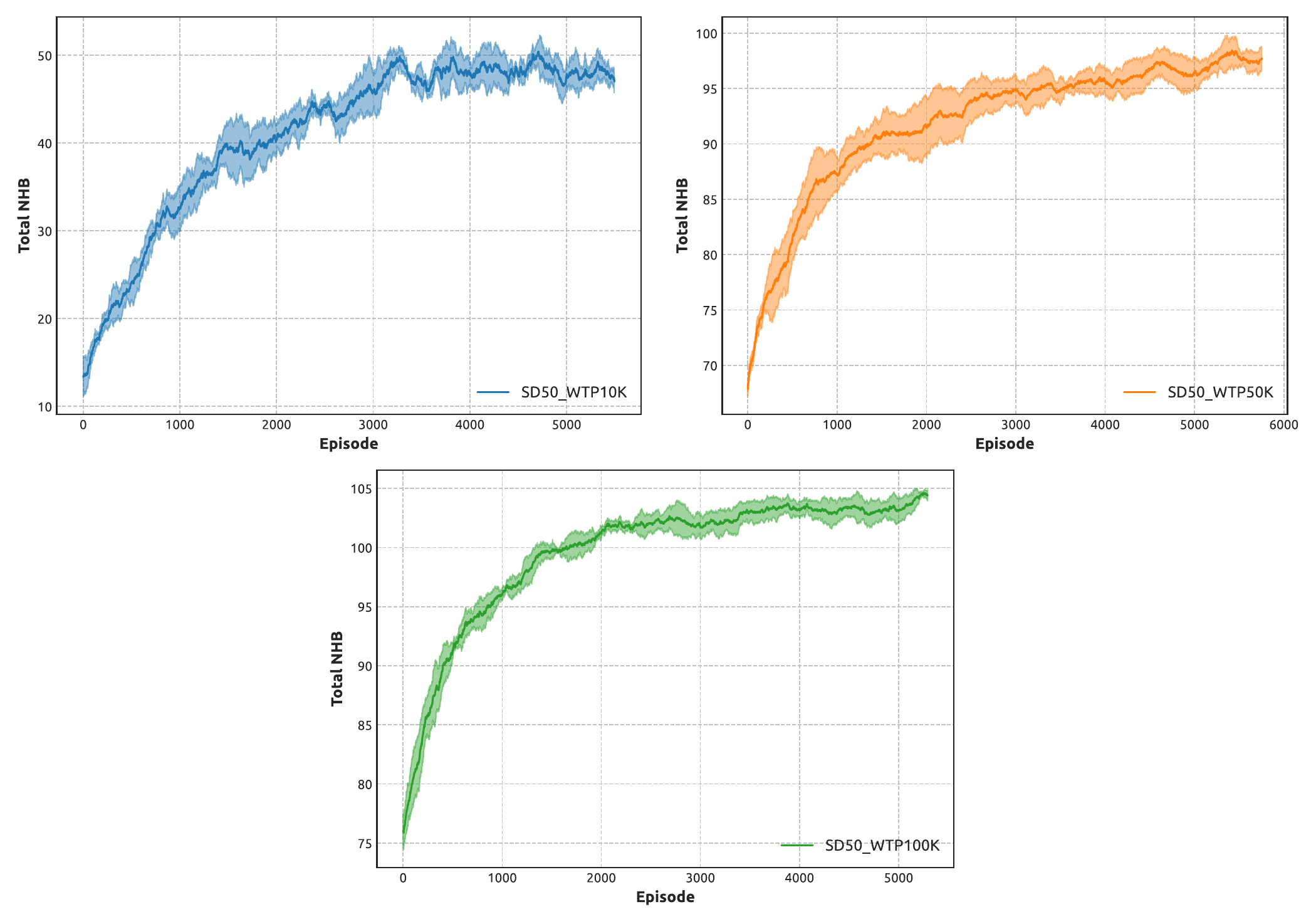}
    \caption{Accumulated episodic rewards generated over training episodes by the SD interventions trained under different combinations of WTP and $SD_{max}=0.5$. Solid curves represent the mean values and the shaded areas represent standard deviation.}
    \label{fig:rewards_training_SD50}
\end{figure}

\begin{figure}[ht]
    \centering
    \includegraphics[width=\textwidth]{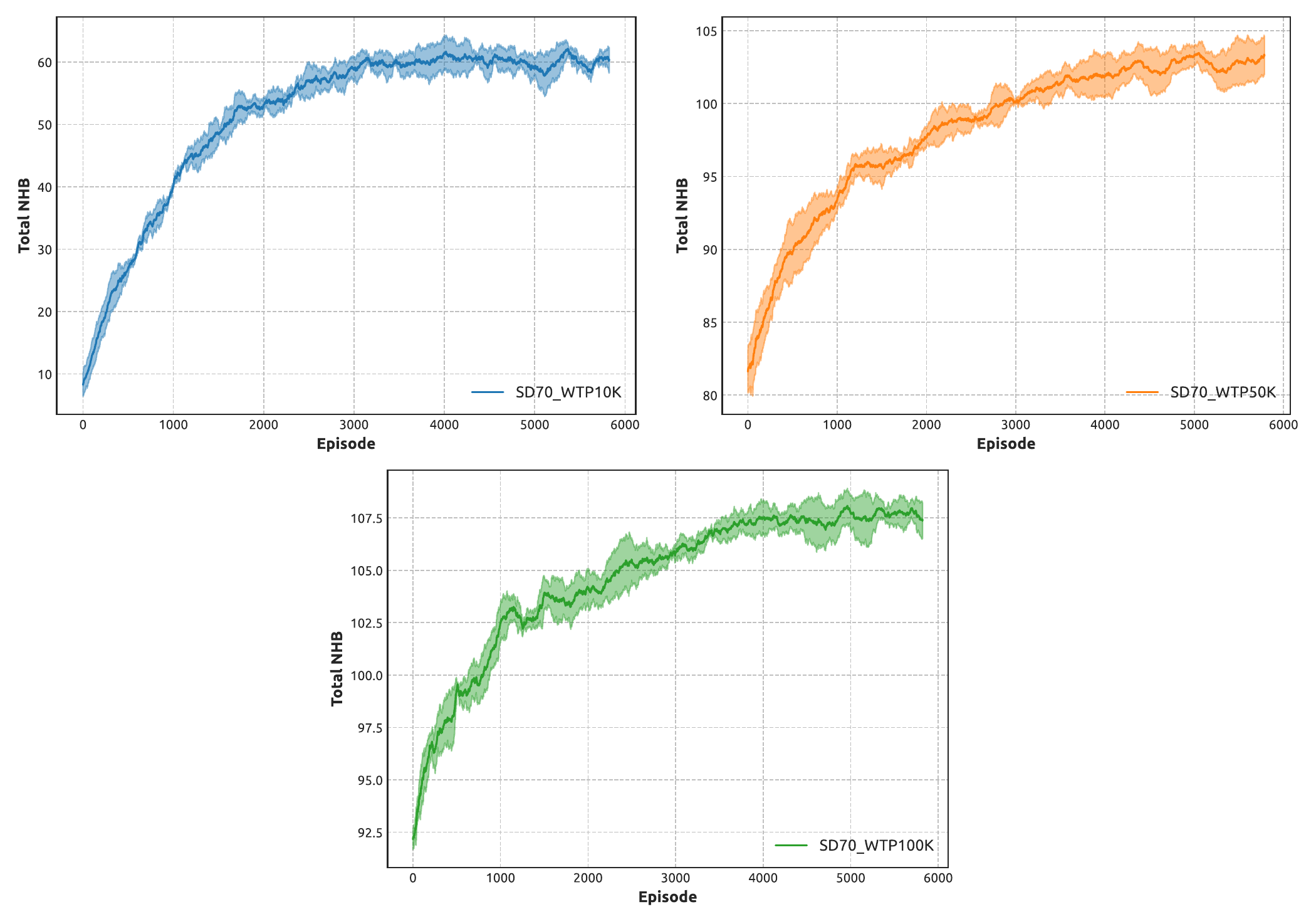}
    \caption{Accumulated episodic rewards generated over training episodes by the SD interventions trained under different combinations of WTP and $SD_{max}=0.7$. Solid curves represent the mean values and the shaded areas represent standard deviation.}
    \label{fig:rewards_training_SD70}
\end{figure}

\subsection{PPO Algorithm}
\label{sec:SM_PPO}
Proximal Policy Optimization (PPO) algorithm was developed by~\cite{schulman_proximal_2017} as a policy gradient actor-critic method which approximates both policy and value function. The architecture includes two components. The first component is the actor neural network used to make decisions about actions. The second component is the critic neural network used to predict the value function which is defined as the expected accumulated reward that the decision-maker can receive since a state. In our PPO implementation, the value function $V(s)$ is approximated by the critic network based on the observation of the decision-maker when the state of the environment is $s$. The critic neural network can be trained using the Temporal Difference (TD) Learning with the loss function specified by $L_t^{Vf} = \delta_t^2$, for 
\begin{ceqn}
\begin{equation}
    \delta_t = r_{t+1} + \gamma V(s_{t+1}) - V(s_{t}) 
\end{equation}
\end{ceqn}
\noindent where $r_{t+1}$ is the reward received at the time step $t+1$, $\gamma$ is the discount factor, and $s_{t}$ and $s_{t+1}$ are the states of the environment at the time steps $t$ and $t+1$ along the epidemic trajectory $\tau$ controlled by the SD intervention. 

The objective function used to optimise the actor is a clipped ``surrogate" objective function defined as follows:
\begin{ceqn}
\begin{equation}
    \label{eq:clipped_PPO_objective}
    J(\theta_k) = L^{CLIP}(\theta_k) = \hat{\mathop{\mathbb{E}}}_t \left[ \text{min} \left( \phi_t(\theta_k)\hat{A}_t, \text{clip} \left(\phi_t(\theta_k), 1-\epsilon, 1+\epsilon \right) \hat{A}_t \right) \right]
\end{equation}
\end{ceqn}
\noindent where $\hat{\mathop{\mathbb{E}}}_t$ denotes the empirical expectation over a batch of samples at the time step $t$, index $k$ is the current learning step, $\theta$ is the set of the parameters determining the actor, $\phi_t(\theta_k) = \frac{\pi_{\theta_k}(a_t | s_{t})}{\pi_{\theta_{k-1}}(a_t | s_{t})}$ is the ratio of the probabilities of taking action $a_t$ according to the stochastic policies parameterised by $\theta_k$ and $\theta_{k-1}$, with $\theta_{k-1}$ being the parameters of the previous policy used before the update, $\epsilon$ is a hyperparameter that limits the range of $\phi_t(\theta_k)$ to $[1-\epsilon, 1+\epsilon]$ within the \textit{clip} operation. Finally, $\hat{A_t}$ is the estimation for the advantage function at the time step $t$ which is computed as follows:
\begin{ceqn}
\begin{equation}
    \label{eq:advantage_function}
    \hat{A}_t = \delta_t + (\gamma \upsilon)\delta_{t+1} + ... + (\gamma \upsilon)^{T-t+1}\delta_{T-1}
\end{equation}
\end{ceqn}
\noindent where $\gamma$ is the discount factor, and $\upsilon$ is the weight discount used in the generalised advantage estimation by the policy gradient implementation. In our study, we applied the standard version of PPO algorithm \cite{schulman_proximal_2017} implemented by \cite{stable-baselines3} with the default set of hyperparameters. 

\subsection{Extended Results}
In this study, we modelled adaptive SD interventions, optimised for their cost-effectiveness across different settings of maximal compliance with social distancing ($SD_{max}$) and ``willingness to pay'' (WTP). Section 2 presented our results and analysis contrasting two $SD_{max}$ levels, 30\% and 70\%, across all three considered WTP levels ( \$10K per DALY, \$50K per DALY, and \$100K per DALY). In this section, we include additional figures, comparing the medium and high $SD_{max}$ settings: 50\% and 70\%. As before, the training and simulations for each value of $SD_{max}$ were repeated for all three WTP levels. The dynamics of the optimised adaptive NPIs (Fig. \ref{fig:eval_policies_SD50_SD70}), the resultant dynamics of net health benefit (Fig. \ref{fig:cumulative_rewards_SD50_SD70}), and the associated NHB components including the economic costs and health effects (Fig. \ref{fig:economic_cost_and_health_effect_SD50_SD70}) are presented to complement figures shown in Section 2.

\begin{figure}[ht]
    \centering
    \includegraphics[width=\textwidth]{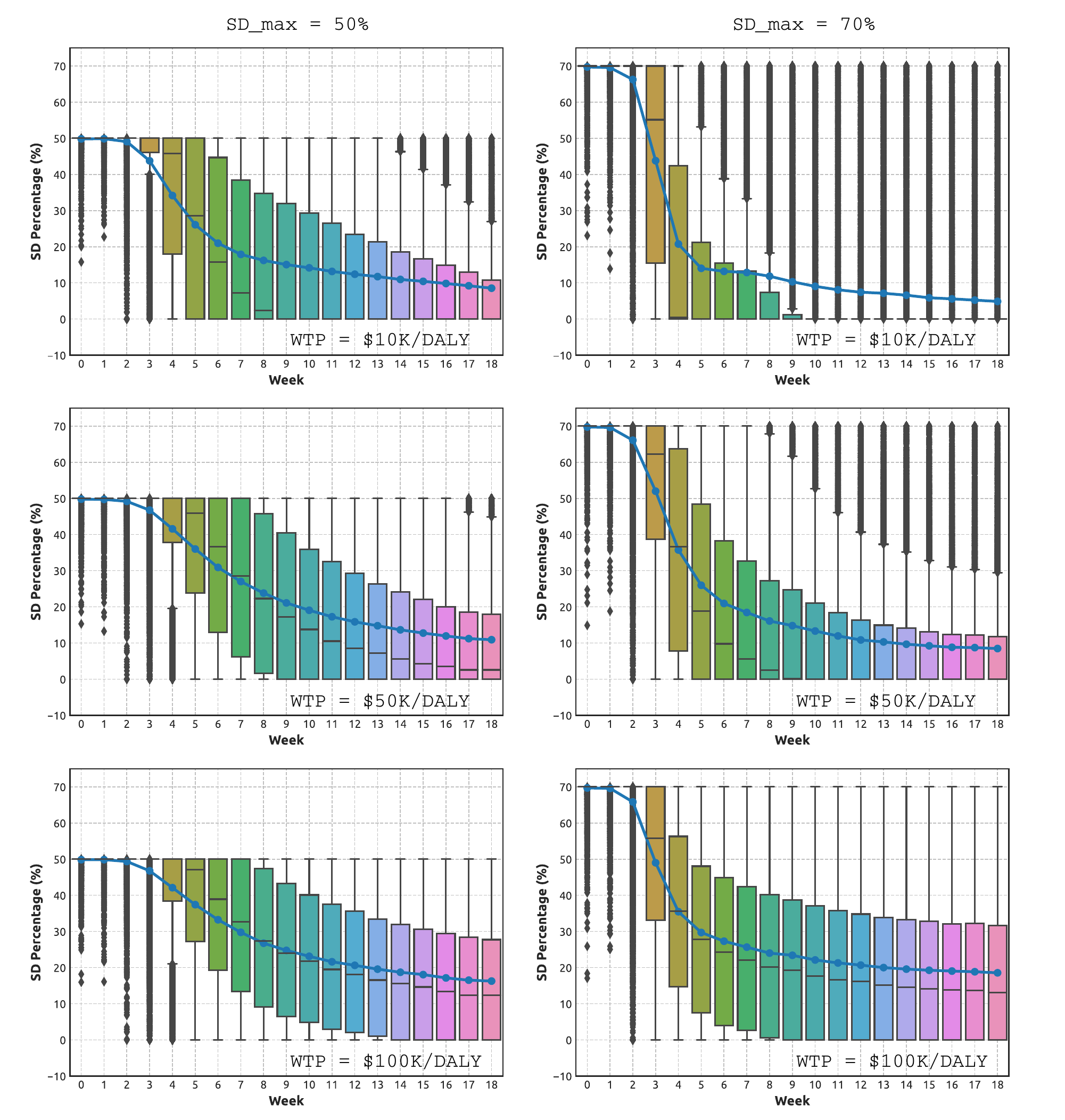}
    \caption{Adaptive NPIs, learned under different combinations of maximal SD levels $SD_{max}$ and WTP, over more than 14,000 simulations. Left: $SD_{max} = 0.5$. Right: $SD_{max} = 0.7$. Top: WTP is set at \$10K per DALY. Middle: WTP is set at \$50K per DALY. Bottom: WTP is set at \$100K per DALY. Boxplots show the distribution of data over the quartiles, with box body capturing the mid-50\% of the distribution. The curves shown with blue colour trace the mean values of the SD levels attained in each week. Outliers are shown in black.}
    \label{fig:eval_policies_SD50_SD70}
\end{figure}

\begin{figure}[ht]
    \centering
    \includegraphics[width=\textwidth]{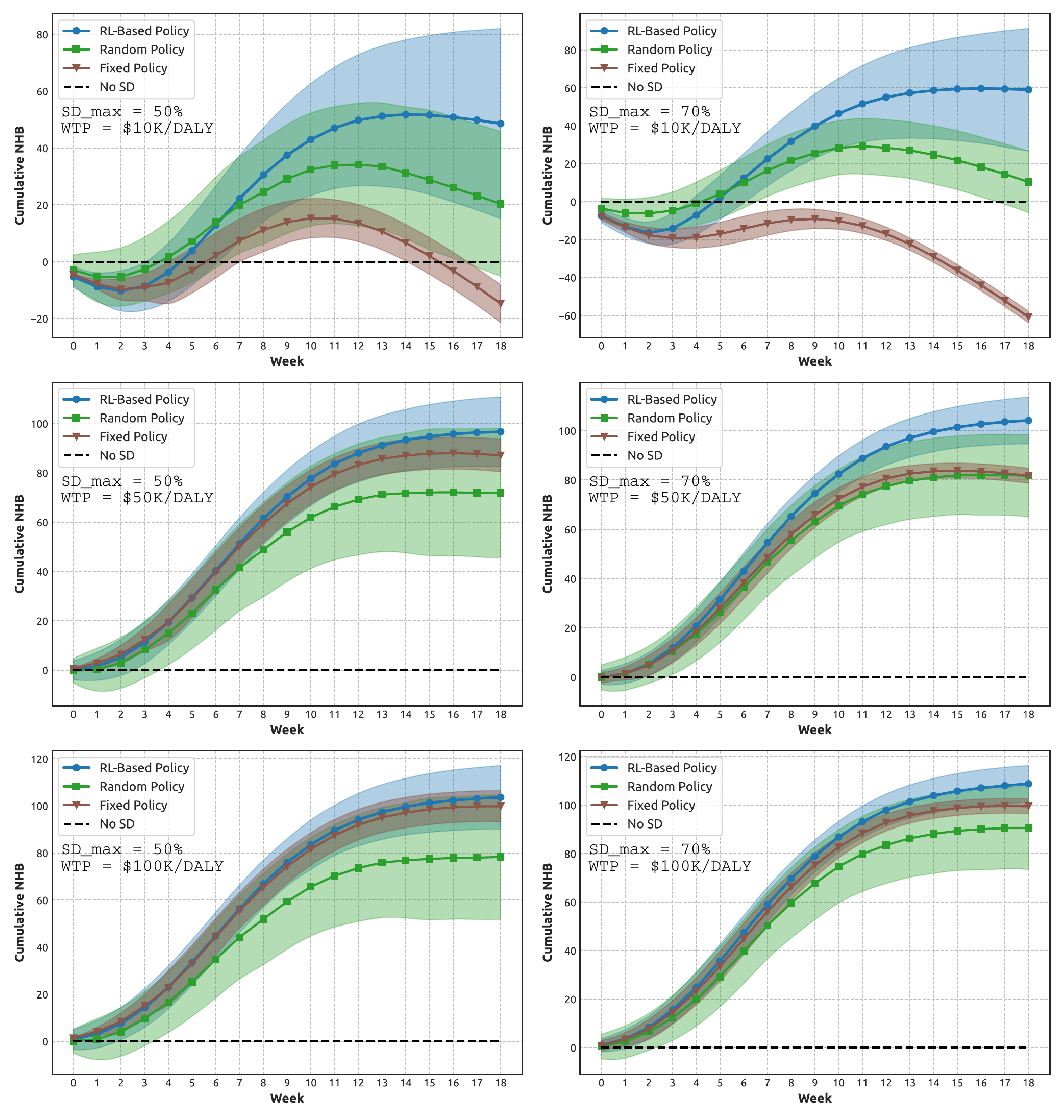}
    \caption{A comparison of cumulative Net health benefit (NHB) generated by the adaptive NPIs, fixed SD NPIs, random SD policies, and zero SD policies.  Left: $SD_{max} = 0.5$. Right: $SD_{max} = 0.7$. Top: WTP is set at \$10K per DALY. Middle: WTP is set at \$50K per DALY. Bottom: WTP is set at \$100K per DALY. Shaded areas show standard deviation.}
    \label{fig:cumulative_rewards_SD50_SD70}
\end{figure}

\begin{figure}[ht]
    \centering
    \includegraphics[width=\textwidth]{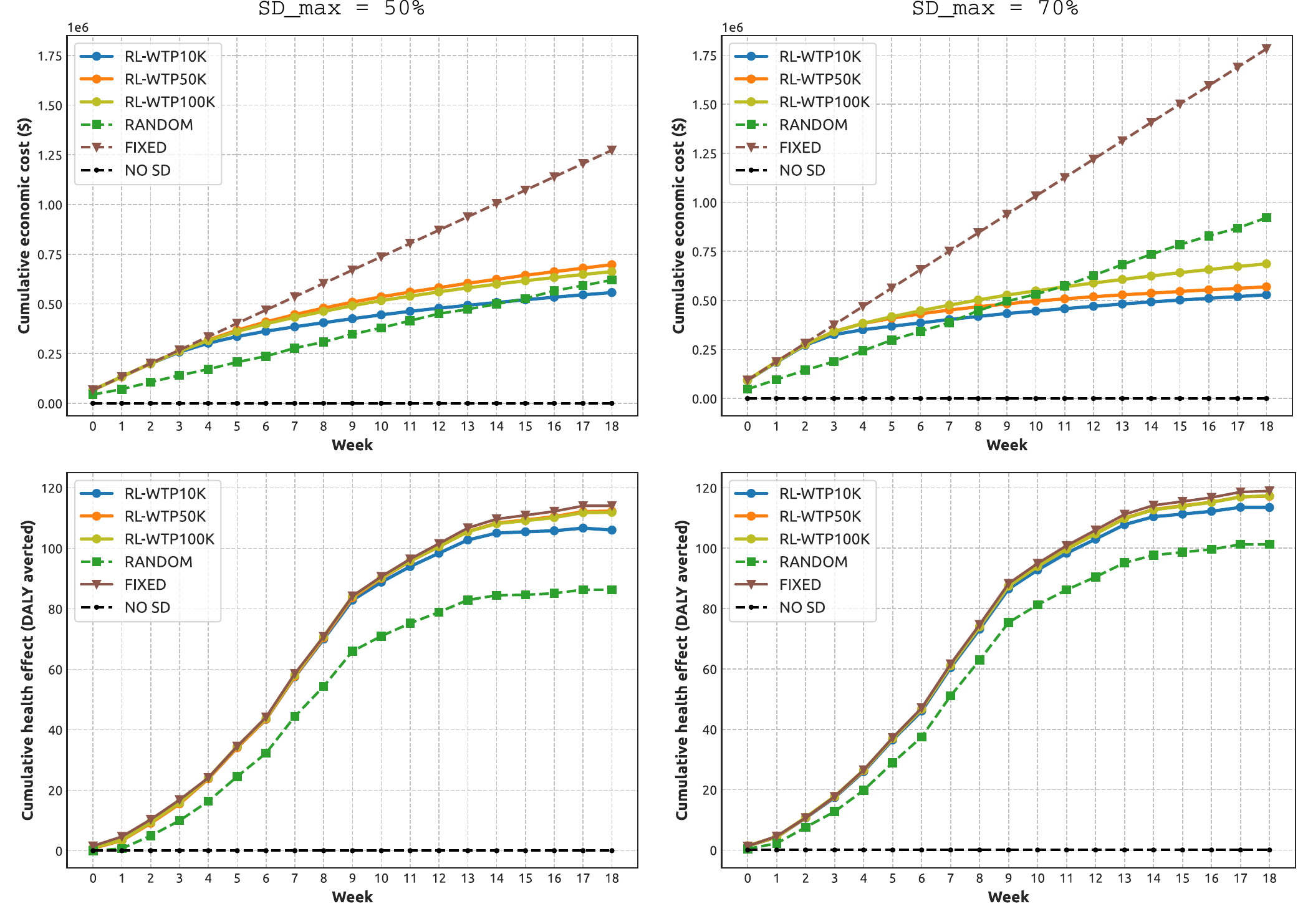}
    \caption{Components of the Net Health Benefit (NHB): mean values of cumulative economic costs (dollars) and cumulative health effect (DALY averted), shown for different NPIs: adaptive SD with three WTP thresholds (\$10K per DALY, \$50K per DALY and \$100K per DALY), random SD, fixed SD, and zero SD.  Left: the maximal SD level $SD_{max}$ is set at 50\%. Right: the maximal SD level $SD_{max}$ is set at 70\%.}
    \label{fig:economic_cost_and_health_effect_SD50_SD70}
\end{figure}

\end{appendices}

\clearpage

\bibliographystyle{unsrt}  
\bibliography{references}

\end{document}